%% file: Vibrations_JSV_V4.tex
\documentclass{elsart}

\usepackage{amssymb}
\usepackage{amsmath}
\usepackage{amsfonts}
\usepackage{bm}

\usepackage{array}
\usepackage{dashrule}
\usepackage{dcolumn}
\usepackage{dsfont}
\usepackage[T1]{fontenc}
\usepackage{longtable}
\usepackage{setspace}
\usepackage{verbatim}

\usepackage[english]{babel}
\usepackage{latexsym}

\usepackage{epsfig}

\usepackage{rotate}
\usepackage{color}
\usepackage{textcomp}
\usepackage{multirow}
\usepackage[section]{placeins}

\newcommand{\CR}{\\}

\usepackage{hyperref}
\newcommand{\figureplace}{}
\newcommand{\tableplace}{}

\newcommand{\Commentaire}[1]{}

\newcommand{\idr}[1]{_{\text{#1}}}

\newcommand{\dd}{\text{d}}
\newcommand{\ee}{\text{e}}

\newcommand{\ED}{\end{document}}

\begin{document}

\begin{frontmatter}

\title{Vibroacoustics of the piano soundboard: (Non)linearity and modal properties in the low- and mid-frequency ranges} 
\author{Kerem Ege\corauthref{cor1}}
\ead{kerem.ege@insa-lyon.fr}
\address{Laboratoire Vibrations Acoustique, INSA-Lyon, 25 bis avenue Jean Capelle, \mbox{F-69621 Villeurbanne Cedex}, France}
\corauth[cor1]{corresponding author}

\author{Xavier Boutillon}
\ead{boutillon@lms.polytechnique.fr}
\address{Laboratoire de Mécanique des Solides (LMS), École polytechnique, \mbox{F-91128 Palaiseau Cedex}, France}

\author{Marc Rébillat}
\ead{marc.rebillat@ensam.eu}
\address{Laboratoire Procédés et Ingénierie en Mécanique et Matériaux (PIMM), Arts et Métiers ParisTech, 151 boulevard de l'Hôpital, \mbox{F-75013 Paris}, France}

\begin{abstract}
The piano soundboard transforms the string vibration into sound and therefore, its vibrations are of primary importance for the sound characteristics of the instrument. An original vibro-acoustical method is presented to isolate the soundboard nonlinearity from that of the exciting device (here: a loudspeaker) and to measure it. The nonlinear part of the soundboard response to an external excitation is quantitatively estimated for the first time, at $\approx -\,40$~dB below the linear part at the \emph{ff} nuance. Given this essentially linear response, a modal identification is performed up to 3~kHz by means of a novel high resolution modal analysis technique~(Ege \emph{et al}., High-resolution modal analysis, JSV, 325(4-5), 2009). Modal dampings (which, so far, were unknown for the piano in this frequency range) are determined in the mid-frequency domain where FFT-based methods fail to evaluate them with an acceptable precision. They turn out to be close to those imposed by wood. A finite-element modelling of the soundboard is also presented. The low-order modal shapes and the comparison between the corresponding experimental and numerical modal frequencies suggest that the boundary conditions can be considered as blocked, except at very low frequencies. The frequency-dependency of the estimated modal densities and the observation of modal shapes reveal two well-separated regimes. Below $\approx$~1~kHz, the soundboard vibrates more or less like a homogeneous plate. Above that limit, the structural waves are confined by ribs, as already noticed by several authors, and localised in restricted areas (one or a few inter-rib spaces), presumably due to a slightly irregular spacing of the ribs across the soundboard.
\end{abstract}


\end{frontmatter}

\section{Introduction}
\begin{figure}[ht!]
\begin{center}
\includegraphics[width=0.8\linewidth]{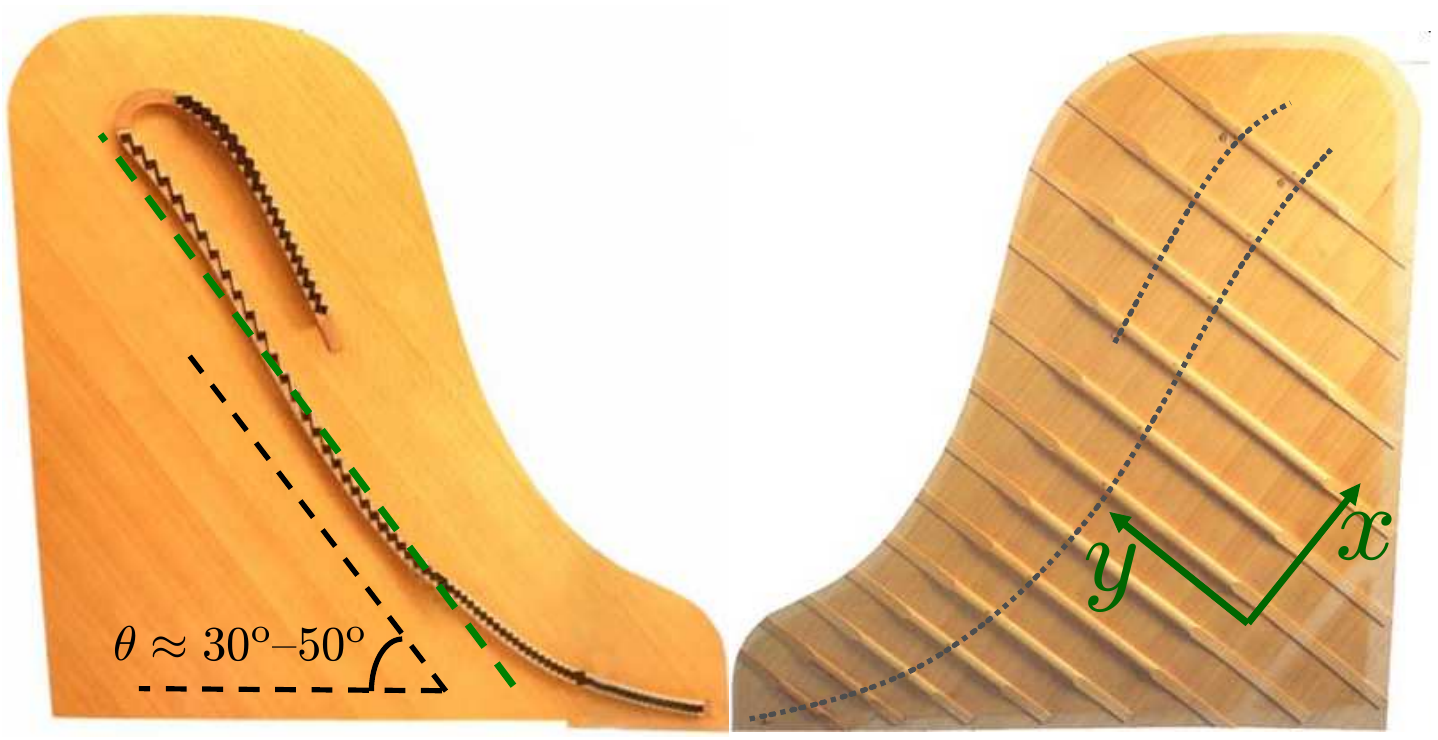}
\caption[Soundboard]{Soundboard of a grand-piano (http://www.lindebladpiano.com). Left: upper face, with bridges (bars where strings are attached) visible. Right: lower face, with ribs. Note that ribs are closer one to each other in the treble range of the instrument (lower part of the picture).
}
\label{fig:board_twofaces}
\end{center}
\end{figure}

Since the strings of a piano are too thin to radiate sound, this function is ensured by the soundboard, a complex plate made of many wooden parts. The sound of the instrument is therefore largely determined by the vibrational characteristics of the soundboard. As shown in Figs.~\ref{fig:board_twofaces} and \ref{fig:table_exp_maillage}, the main element is a large and thin panel, made out of glued strips (usually in spruce). We define the $x$-direction as the grain direction of this panel's wood. In the $y$-direction are glued a series of parallel, nearly equidistant stiffeners, called ribs and also made out of spruce (sometimes sugar pine). On the opposite face are glued the bridges: one short and one long thick bars running approximately in the $x$-direction, slightly curved, made out of maple, on which the strings are attached. The overall shape of this ensemble depends on the type of piano: nearly rectangular for upright pianos, the shape of a half round-and-high hat for grand pianos. The width of the soundboard is more or less 140~cm, corresponding to that of the keyboard. The height or length ranges from more or less 60~cm for small uprights to more than 2~m for some concert grands. The panel thickness $h$ is between 6 and 10~mm, the inter-rib distance $p$ ranges from 10 to 18~cm (depending on pianos). The treble bridge is usually 3 to 4~cm high and 2.5 to 3.5~cm wide. The bass bridge is about 2-3~cm higher than the treble one so that the bass strings can be strung over the lower mid-range ones. The dimensions of the cross section of the ribs are about $25$~mm in width with height varying from $15$ to $25$~mm (thinner toward the treble region).

The functioning of the piano goes schematically as follows: once a key has been struck, a hammer strikes one, two, or three tuned strings and goes back to its rest position, leaving the strings vibrating freely. The bridge represents a nearly fixed end for the strings so that energy stored initially in the strings is slowly transferred from the strings to the soundboard (a significant quantity of energy being also dissipated inside the strings themselves), partly dissipated inside wood and partly radiated acoustically. Since the decay time is several orders of magnitude larger than the periods of vibration, it can be considered that the soundboard is put into a forced motion by the strings, at frequencies that have no relationship with the resonance frequencies of the soundboard itself. Since the string and the soundboard can be considered as  almost dynamically uncoupled, it makes sense to analyse the dynamics of the soundboard in terms of its normal modes.

The present article is focused on some main features of the vibration regimes of the soundboard of an upright piano, as observed in playing condition: linearity, modal density and modal damping. The literature on the vibrations of the soundboard has been recently reviewed in~\cite{EGE2009_2bib}. Curiously, the literature is almost mute on the first point, even though linearity is a requisite for the usual representations of the dynamics of the piano soundboard -- modes, mechanical impedance or mobility -- to be physically meaningful quantities. Section~\ref{sec:Linearity} presents the first, to the best of our knowledge, quantitative evaluation of the (non)linearity of the soundboard vibration. Modal analyses are presented next: experimental in Sec.~\ref{sec:Method} and numerical, by means of a finite-element modelling of the soundboard in Sec.~\ref{sec:FEM}. The frequency range under scope in this article \mbox{[0 - 3] kHz} is much wider than in most previous experimental studies devoted to modal parameters of the piano. This could be achieved thanks to the use of a recently published high-resolution modal analysis technique~\cite{EGE2009}. Compared to techniques based on the Fourier transform, it avoids the customary compromise in time-frequency resolution and thus, gives access to an extended frequency-range. Since the timbre of piano depends highly on the relative decay-times of the components of each note, energy dissipation represents an important dynamical parameter which is accessed here \emph{via} the modal dampings of the soundboard. The evaluation of modal dampings could be performed here up to several kHz for the first time in piano studies (Sec.~\ref{sec:dampings}). The results are presented and discussed in Sec.~\ref{sec:Results} in terms of modal shapes in the low-frequency range and in terms of modal densities and dampings up to 3~kHz. Analyses are done in the spectral domain throughout the whole paper and physical quantities are thus considered as implicit or explicit functions of the frequency $f$.

\input{Linearity_V41}

\input{ESPRIT_V42}

\input{FEM_V42}

\input{Results_V43}

\section{Conclusion}
We have applied original techniques to investigate the vibrations of the soundboard of an upright piano in playing condition. The nonlinear part of the mechanical response to an acoustical excitation could be separated from the nonlinear contribution induced by the loudspeaker. At levels of vibration corresponding to \emph{ff} playing, the nonlinear component of the soundboard vibration is $\approx$~30--50 dB below the linear part, in the \mbox{[100 - 3500] Hz} frequency range. It is likely that the main nonlinearity is a consequence of large displacements of the soundboard (geometric nonlinearity). If this is true, vibrations at higher frequency and corresponding to comparable acoustical levels are not likely to generate larger nonlinear components. One may therefore safely retain the order of magnitude of -40 dB for the distortion rate at the \emph{ff} nuance, for the piano that has been investigated. For larger pianos, such nonlinearities might be expected to be even less. This preliminary study shows that a linear model is sufficient to predict the main features of the vibro-acoustical behaviour of a piano soundboard in playing situations.

Given the essentially linear response, modal identifications have been performed between 50~Hz and 3 kHz by means of a novel high-resolution modal analysis technique. For the piano, this frequency range belongs mostly to the mid-frequency domain since the modal overlap appears to range from 30\% at 150 Hz to 100\% at 1 kHz, decreasing down to 60\% at 3 kHz. The loss factor appears to be maintained between 1 and 3\% over several kHz, with a significant dispersion but without strong systematic variations. The dispersion might be attributed to the different acoustical efficiencies of the different modes of the soundboard. Since the loss-factor commonly observed for spruce is about 2\%, the energy dissipation scheme is likely to be that in which only a small part of the power is radiated acoustically at any frequency, thus providing a long decay time for each note, at each of its partial frequencies (commonly but incorrectly called "harmonics"). On one hand, this raises the hope that using a less lossy material than wood may provide a higher sound level together with keeping the same decay-rate for each note. However, two other requirements for a good tonal quality of the instrument must be kept in mind: spectrum regularity (one partial must not behave too differently from the other ones) and homogeneity along the tonal range (notes must not differ appreciably, at least from their neighbours in pitch). Since the efficiency of the acoustical radiation of the soundboard is expected to be much more frequency-dependent than loss-factors in wood, energy losses that would be caused primarily by acoustical radiation must be expected to affect negatively spectrum regularity and tonal homogeneity. In other words, the fact that in today's pianos, losses seem to be mainly located in wood certainly smoothens frequency-dependency of the decay rates, which is favourable for these two timbre qualities.

The frequency evolution of the estimated modal density of the piano soundboard reveals two well-separated vibratory regimes of the structure. Below approximately 1~kHz, the modal density and the modal shapes look like those of a homogeneous plate. The vibration extends over the whole area of the soundboard, including in the so-called "dead-zones". Analysing together the modal shapes, the modal frequency and the evolution of the modal density in the low-frequency domain suggests that boundary conditions can be considered as (a) ruled by inertia for the one or two very low modes and (b) constrained in general. The scheme that we propose for boundary conditions is that the rotational degrees of freedom are blocked whereas the translational degrees of freedom are massive.

Above 1~kHz, the soundboard operates as a set of structural wave-guides defined by the ribs, as already noticed by several authors. The modal shapes obtained by FE-modelling confirm this confinement and suggest that modes are \emph{localised} in restricted areas (one or a few inter-rib spaces), due to a slightly irregular spacing of the ribs across the soundboard.

All these observations pave the way for a very synthetic modelling of the soundboard vibration up to several kHz.

\section*{Acknowledgements}
This work has been initiated during the PhDs of the first and third authors at the LMS, for which they were sponsored by the French Ministry of Research. We express our gratitude to Andrei Constantinescu for his precious help on the implementation of the FEM of the piano soundboard. We thank Stephen Paulello for sharing his knowledge of piano making with us on a number of questions.

\begin{appendix}

\section{\appendixname: Volterra series for modelling weakly nonlinear systems}
\label{app:Volterra}
\input{Appendix_MR}

\section{\appendixname: Estimation of the kernels of a cascade of Hammerstein models}
\label{app:NL_method}
\input{NL_Method}

\section{\appendixname: Dimensions of the ribs for the FEM}
\label{sec:RibDimensions}
\input{appendix_ribsV32}
\end{appendix}
\clearpage\newpage

\bibliographystyle{elsart-num}
\bibliography{SoundboardVibrationV35}

\end{document}

%% file: Linearity_V41.tex
\section{A linear behaviour ?}
As recalled above, linearity of the piano must be guaranteed if the usual representations (modes, impedance) are to be used. Moreover, people have expressed the opinion that nonlinearities could be significant in the sound of the instrument. Both questions reduce to a quantitative aspect (how small?) but with the same reference: a 1\% distortion rate (-40 dB) may be considered as sufficient small for using linear physical concepts but would be considered as mediocre for audio equipment.

\label{sec:Linearity}
\subsection{Past studies}
\begin{figure}[ht!]
\begin{center}
\includegraphics[width=0.7\linewidth]{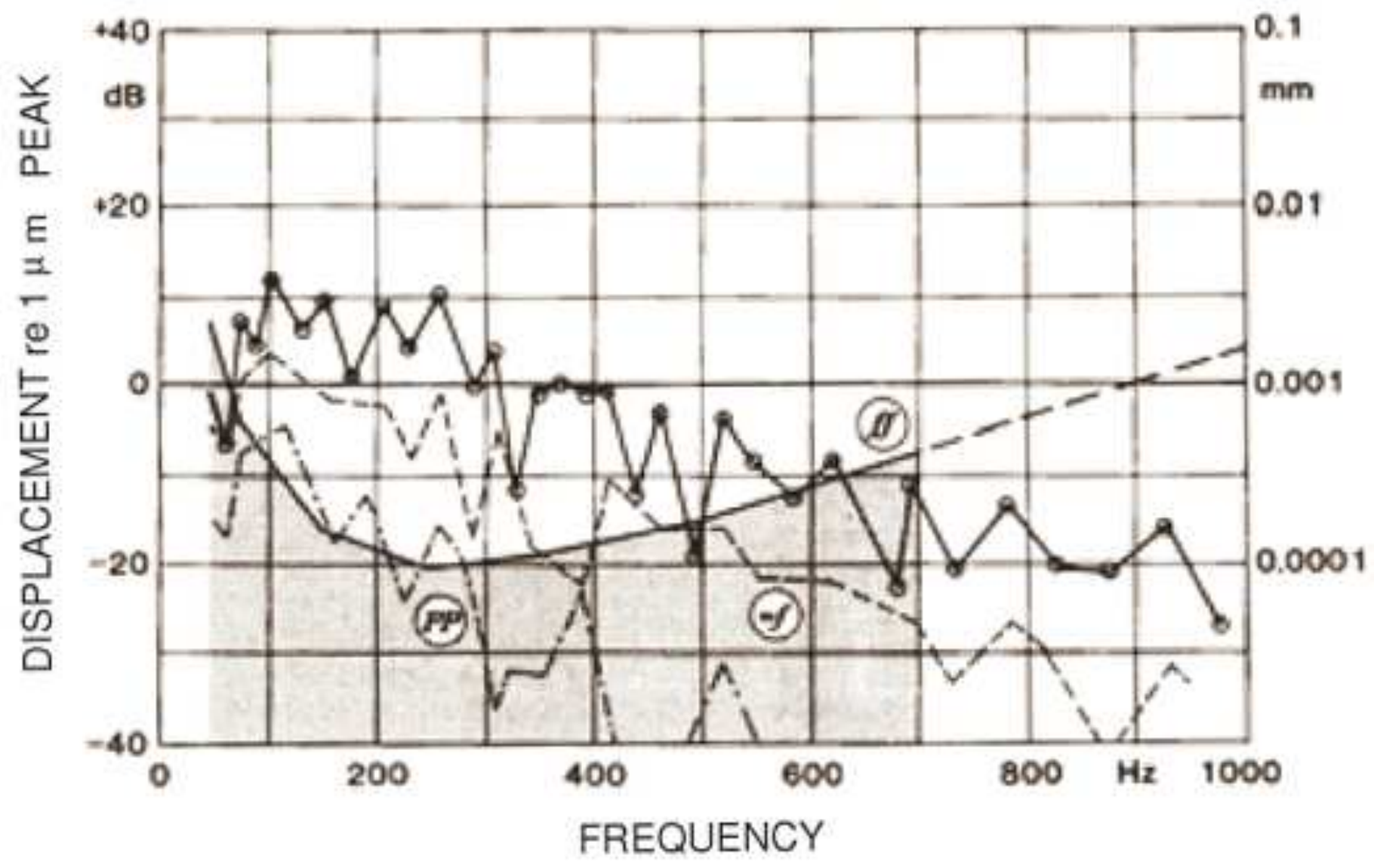}
\end{center}
\caption[Ask]{Vibration levels at the bridge of a grand piano when played \emph{pp} (dash-dotted line), \emph{mf} (dotted line) and \emph{ff} (solid line with $\bullet$ marks) for the notes $\mathbf{C_2}$ to $\mathbf{B_5}$ (fundamental frequencies $\approx$~60 to 950~Hz), according to Askenfelt and Jansson~\cite{ASK1992}. The region below the threshold of vibration sensation at the fingertips (contact area 28~mm$^2$) is shown shaded.}
\label{fig:Ask_soundb_disp}
\end{figure}

Nonlinear phenomena such as jump phenomenon, hysteresis or internal resonance appear when the transverse vibration of a bi-dimensional structure exceeds amplitudes in the order of magnitude of its thickness~\cite{TOU2002}. In the case of the piano, the soundboard transverse motion measured at the bridge remains in a smaller range, even when the piano is played \emph{ff} in the lower side of the keyboard. Askenfelt and Jansson~\cite{ASK1992} report maximum values of the displacement at the bridge of \mbox{$\approx~6\cdot 10^{-6}$~m} in the frequency range \mbox{[80-300]~Hz} (Fig.~\ref{fig:Ask_soundb_disp}). This maximum value is less than $10^{-3}$ times the board thickness. We can therefore expect that, to a high level of approximation, the vibration of the soundboard is linear.

\renewcommand{\figureplace}{%
\begin{center}
[\figurename~\thepostfig\ about here (with ref.~\cite{ASK1992} in caption).]
\end{center}}

\renewcommand{\figureplace}{%
\begin{center}
[\figurename~\thepostfig\ about here.]
\end{center}}

\begin{figure}[ht!]
\begin{center}
\includegraphics[width=1\linewidth]{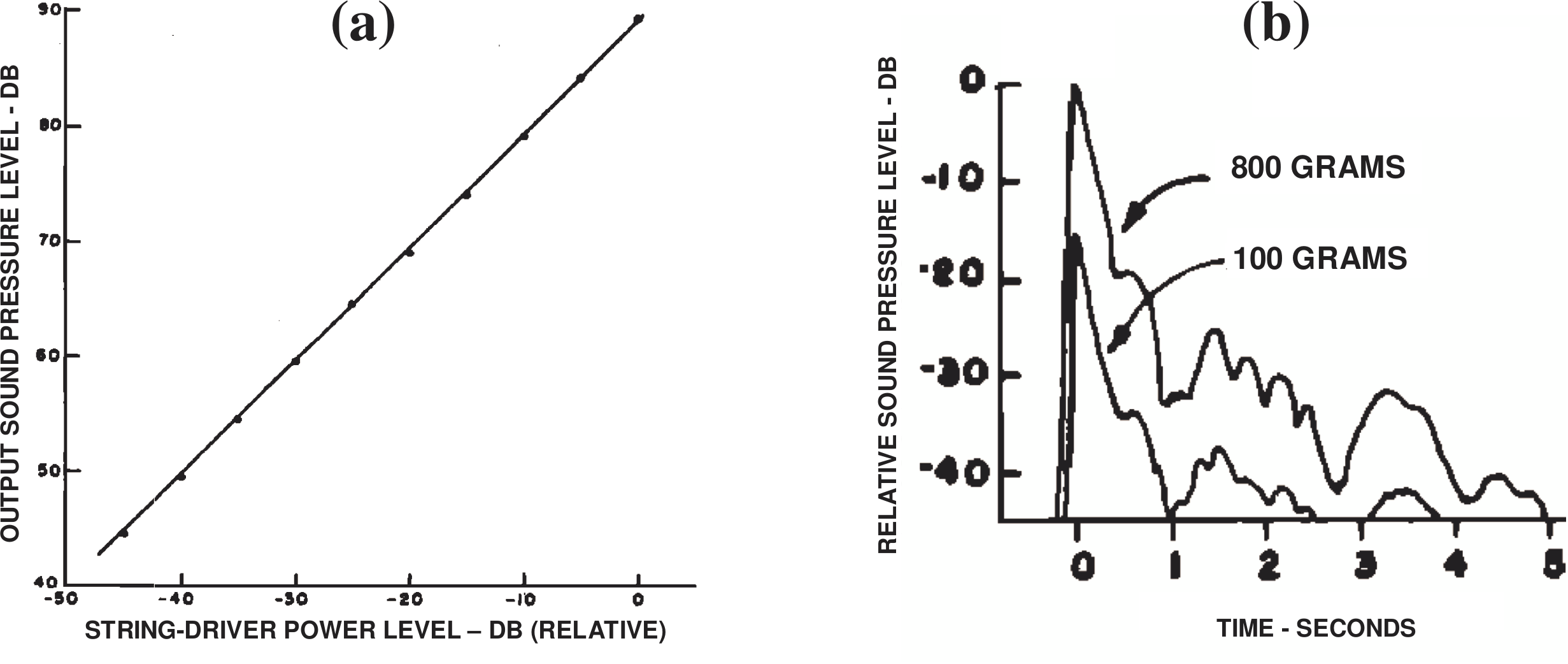}
\end{center}
\caption[Hundley]{Linearity of response for the \emph{string-to-bridge-to-soundboard-to-air path}, according to Hundley \textit{et al.}~\cite{HUN1978}. (a) with an electromagnetically excited string (after Fig.~5). (b) with different weights actuating the C$_5$ key (after Fig.~6).}
\label{fig:Hundley_linear}
\end{figure}

Very few experimental studies have been carried out concerning the linearity of the piano soundboard vibrations. The most convincing work seems to be that by Hundley \textit{et al.}~\cite{HUN1978}. Their motivation was in fact to eliminate nonlinearity as one possible cause of the multiple time-decay of piano tones. Therefore, these authors studied the nonlinearity in the so-called \emph{string-to-bridge-to-soundboard-to-air path}. A direct measure of linearity was obtained by using a magnetic driver close to a string triplet (note $\mathbf{B_3}$) and driving the string(s) at a fundamental frequency of 250~Hz. The curve shown in Fig.~\ref{fig:Hundley_linear}(a) represents the sound pressure level near the piano in an anechoic environment for different input levels of the driver. Proportionality between the sound pressure level and the excitation level is excellent up to a SPL of 90 dB, with the slope of the curve being very close to one ($\approx0.995$). However, our opinion is that no quantitative evaluation of the nonlinearity can be withdrawn from this curve. Since the main response is linear, nonlinearities manifest themselves only at high vibration levels so that the major part of this curve do not carry useful information. In order to extract an order-of-magnitude for the nonlinear part, the experimental precision at the highest point must be at least one order of magnitude higher: 60 dB for measuring a 1\% distortion, for example. This is clearly not the case in this measurement.

In a second step, the authors compared the tone level recordings for which the key was actuated by different weights (from 100 grams to 800 grams). This experiment was repeated for a large number of keys and no evidence of dependence between the decay rate and the blow force was found: the curves were almost identical (see Fig.\ref{fig:Hundley_linear}(b) for an example of this measurement on note \textbf{C5}). Contrary to the first one, this second experiment is qualitative in nature but yields here a null result.

\renewcommand{\figureplace}{%
\begin{center}
[\figurename~\thepostfig\ about here (with ref.~\cite{HUN1978} in caption).]
\end{center}}

\renewcommand{\figureplace}{%
\begin{center}
[\figurename~\thepostfig\ about here.]
\end{center}}

\subsection{Method to estimate the piano soundboard nonlinearities}
\begin{figure}[ht!]
\begin{center}
\includegraphics[width=0.5\linewidth]{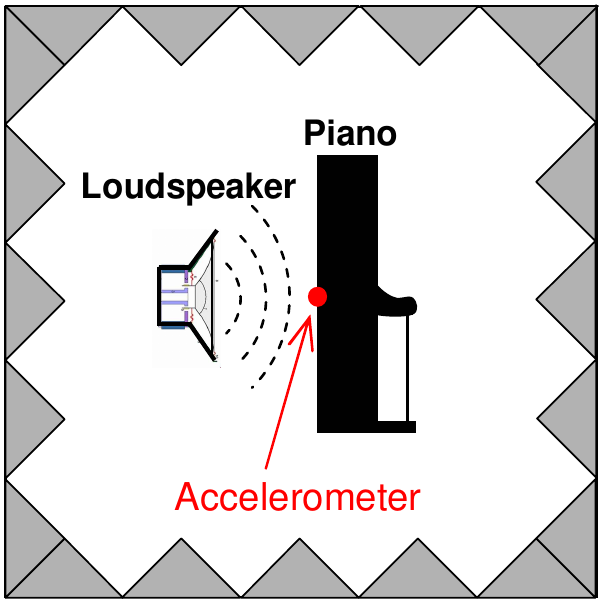}
\end{center}
\caption[aaa]{Acoustical excitation of the piano placed in a pseudo-anechoic room. The acceleration of the board is measured at points A$_1$, A$_2$ and A$_5$ shown in Fig.~\ref{fig:table_exp_maillage}.}
\label{fig:piano_chambre}
\end{figure}

We aim at estimating experimentally the order of magnitude for the ratio between the nonlinear and the linear parts of the soundboard response -- or distortion rate -- when it is excited by a given string force on the bridge. Since replacing the string excitation by a mechanical shaker raises all sorts of experimental problems (attachment, feedback of the piano response on the shaker excitation, control of the side forces, etc.), we have preferred to excite the piano acoustically (with a very powerful electrodynamic loudspeaker) and to evaluate its vibratory response, as shown in Fig.~\ref{fig:piano_chambre}.

We assume that, for a given vibratory level of the soundboard, the distortion ratio is the same, at least in order of magnitude, whether the piano is excited acoustically or mechanically. If the piano soundboard were a point, the former and the latter situations would represent reciprocal experiments and the assumption would be exactly true.

\begin{figure}[ht!]
\begin{center}
\includegraphics[width = 0.7\textwidth]{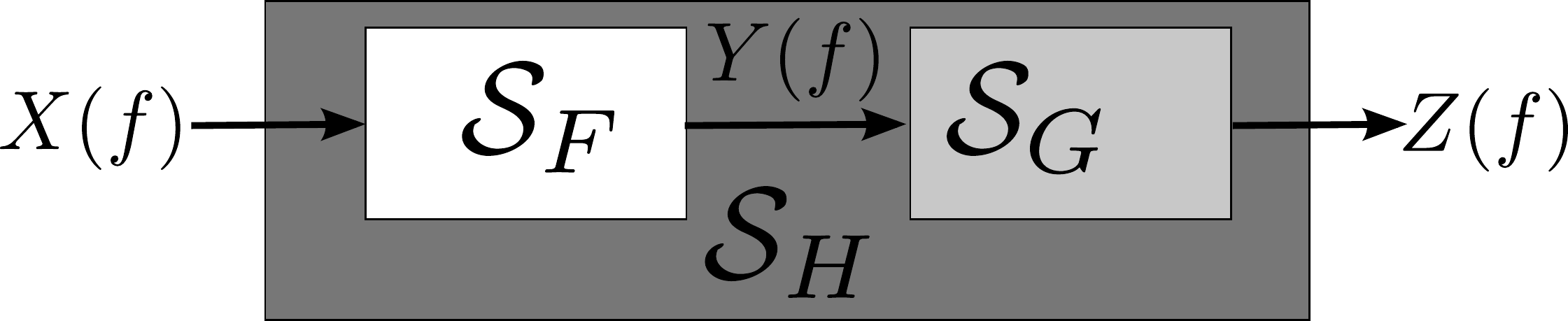}
\caption{A chain of two nonlinear systems models the piano soundboard $\mathcal{S}_G$ excited by the acoustical field created by an electrodynamic loudspeaker $\mathcal{S}_F$. The electrical driving signal is $X(f)$, $Y(f)$ is the acoustical signal (see text for discussion) and $Z(f)$ is the piano vibratory response.}
\label{fig:NL_chain}
\end{center}
\end{figure}

Whether the electrodynamic exciter is acoustical (loudspeaker) or mechanical (shaker), its own nonlinear contribution to the overall response cannot in general be neglected, compared to that of the piano. If the driver response is independent from the piano response (which is true to a large extent in the acoustical case, for a sufficiently powerful loudspeaker), the (free-of-feedback) situation can be represented as in Fig.~\ref{fig:NL_chain}, where the loudspeaker is represented by the system $\mathcal{S}_F$, the piano soundboard by $\mathcal{S}_G$ and their association by $\mathcal{S}_H$. Experimentally, only the input electrical signal of the loudspeaker $X(f)$ can be considered as fully controlled. Since both the exciter and the soundboard are expected to be slightly nonlinear, a direct characterisation of the piano soundboard appears to be very difficult. The method which is presented here derives the distortion rate of $\mathcal{S}_G$ from measurements performed on $\mathcal{S}_F$ and $\mathcal{S}_H$. 

The signal $Z(f)$ denotes the acceleration of the soundboard, possibly at various different locations. For the sake of nonlinearity estimation, it is assumed that the acoustical field created by the loudspeaker can be represented by a scalar value $Y(f)$. Moreover, it is assumed that $Y(f)$ can be estimated by removing the piano and setting a microphone at the place of the soundboard. Again, the numerous corresponding approximations (change of the acoustical field with or without the piano, difference between the acoustical field and its measurement in one point, etc.) are supposed to be correct only as far as the order of magnitude of the nonlinearities is concerned. In other words, we consider that the piano, equipped with accelerometers, behaves, as far as nonlinearities are concerned, like a slightly nonlinear (and localised) microphone. It is explained below how $\mathcal{S}_F$ and $\mathcal{S}_H$ have been characterised and how one can derive the distortion rate of $\mathcal{S}_G$ (the piano soundboard).

The outputs of the systems $\mathcal{S}_F$, $\mathcal{S}_G$ and $\mathcal{S}_H$ can generally be decomposed in their linear and nonlinear parts as follows:			
\begin{align}
\mathcal{S}_F: \quad Y(f)&\stackrel{\Delta}{=} F(f) X(f) [1+C_{\mathcal{S}_F}(f)] 
	\label{eq:def_NL_ESF} \\
	\mathcal{S}_G: \quad Z(f) &\stackrel{\Delta}{=} G(f) Y(f) [1+C_{\mathcal{S}_G}(f)]
	\label{eq:def_NL_ESG}	\\
	\mathcal{S}_H: \quad Z(f) &\stackrel{\Delta}{=} H(f) X(f)[1+C_{\mathcal{S}_H}(f)]
	\label{eq:def_NL_ESH}
\end{align}
where $C_{\mathcal{S}_{F}}(f)$, $C_{\mathcal{S}_{G}}(f)$ and $C_{\mathcal{S}_{H}}(f)$ are the distortion rates of $\mathcal{S}_F$, $\mathcal{S}_G$ and $\mathcal{S}_H$ respectively.

It is assumed that the nonlinearities are mathematically weak: the relationship between their output $s(t)$ and their input $e(t)$ can be expressed analytically by Volterra series~\cite{Palm1978,Boyd1985} (see Appendix~\ref{app:Volterra}). Let the Volterra kernels $\{F_k(f_1,{\scriptstyle \ldots},f_k)\}_{k \in \mathbb{N}^*}$, $\{G_k(f_1,{\scriptstyle \ldots},f_k)\}_{k \in \mathbb{N}^*}$ and $\{H_k(f_1,{\scriptstyle \ldots}, f_k)\}_{k \in \mathbb{N}^*}$ describe the systems $\mathcal{S}_F$, $\mathcal{S}_G$ et, $\mathcal{S}_H$. It is shown in Appendix~\ref{app:Volterra} that, in the chain-case presented in Fig.~\ref{fig:NL_chain}, the Volterra kernels of $\mathcal{S}_H$ are given analytically by functions of the Volterra kernels of $\mathcal{S}_F$ and $\mathcal{S}_G$. For $k=1$, one obtains:

\begin{align}
F_1(f_1) = F(f_1)\qquad G_1(f_1) &= G(f_1)\qquad H_1(f_1) = H(f_1)\\
			H(f) &=F(f) G(f) 
			\label{eq:NL_H1}
\end{align}

This result can be extended to a chain of more than two systems and proves the intuitive result that the linear part of the response of a chain of weakly nonlinear systems is the product of the linear parts (transfer functions) of each system composing the chain.

One assumes further that $\mathcal{S}_F$ and $\mathcal{S}_H$ can be modelled as cascades of Hammerstein models. Cascade of Hammerstein models constitute an interesting subclass of Volterra systems for which parameters can be estimated experimentally~\cite{REB2011,NOV2010}, as explained in Appendix~\ref{app:NL_method}. The purpose here is to derive an estimation of $\mathcal{S}_G$ (the piano soundboard excited by the acoustical field) based on experimental estimations of $\mathcal{S}_F$ (the loudspeaker excited by the driving electrical signal) and $\mathcal{S}_H$ (the piano combined with the loudspeaker, excited by the electrical driving signal).

Using Eqs.~(\ref{eq:NL_H1}),~(\ref{eq:def_NL_ESG}) and multiplying by $F_1(f)$, the distortion rate $C_{\mathcal{S}_G}(f)$ of the piano is given by:
\begin{equation}
	C_{\mathcal{S}_G}(f) = \frac{F(f)Z(f) - H(f) Y(f)}{H(f)Y(f)}
\label{eq:apport_NL_estG}
\end{equation}
where all quantities in the second member of Eq.~(\ref{eq:apport_NL_estG}) can be measured as described in the next section. This method solves the problem of isolating the nonlinearities of a system from those of its exciter.

\subsection{Experimental implementation and results}
An upright piano (Atlas brand) with a rectangular soundboard (dimensions: 0.91~m $\times$ \mbox{1.39 m} $\times$ 8~mm) was used for the experiments. Since the geometries of piano soundboards are rather similar (thickness of the wood panels, height of the rib, width of the soundboard, width/length of the soundboard varying in a 1 to 2 ratio between grands and uprights), the result obtained here can be expected to be similar, in order of magnitude, on other pianos. The piano was put in a pseudo-anechoic room (anechoic walls and ceiling, ordinary ground). It was tuned normally, with strings muted by strips of foam (or woven in two or three places) between them. The electrical excitation $X(f)$ of the loudspeaker (Bose -- 802 Series II) was an exponential swept-sine [50-4000]~Hz (40 kHz sampling frequency, $T=26~\text{s}$ duration). The amplitude of the loudspeaker was adjusted so that the displacement of the soundboard did correspond to the \textit{ff} playing level: $\approx 10^{-6}~\text{m}$ at $\approx370$~Hz in this case.

The acoustic response $Y(f)$ was measured by a microphone (pre-polarised pressure-field 1/2'' -- Brüel \& Kj\ae r 4947) taking place of the piano. It exhibits some distortion which may safely be attributed to the loudspeaker rather than to the microphone (the typical distortion rate of the microphone is 3\% at 160~dB SPL, it becomes totally negligible at a SPL less than 100~dB, compared to that of a loudspeaker, around 2-3\% at 100~dB).
      
The motion $Z(f)$ of the soundboard was measured with three accelerometers (Brüel \& Kj\ae r 4393) put at the locations marked by A$_1$, A$_2$ and A$_5$ in Fig.~\ref{fig:table_exp_maillage}.

The distortion rate $C_{\mathcal{S}_H}(f)$ appears to be generally slightly larger than $C_{\mathcal{S}_F}(f)$. This is consistent with presumably small soundboard nonlinearities and with the expectation that nonlinearities do not compensate each other: nonlinearity is expected to \emph{increase} along the chain. It was also observed that the second order distortion is significantly larger than the third-order distortion and much larger than the other orders, for both $\mathcal{S}_F$ and $\mathcal{S}_H$. In other words, the second-order (and principal) nonlinearity of the loudspeaker seems to extend to the overall "loudspeaker-soundboard" system. Since the soundboard is nearly a flat structure, its intrinsic nonlinearity is expected to be geometric, and of third order. However, the present method does not allow to identify the different nonlinearity orders in $C_{\mathcal{S}_G}(f)$.

In some restricted frequency ranges and for points A$_2$ and A$_5$, it was observed that $C_{\mathcal{S}_F}(f)$ was slightly larger than $C_{\mathcal{S}_H}(f)$. Presumably, occurrences of this anomalous situation can be attributed to the failure of the hypothesis that the piano behaves like a localised system.

\begin{figure}[th!]
\begin{center}
	\includegraphics[width=0.8\columnwidth]{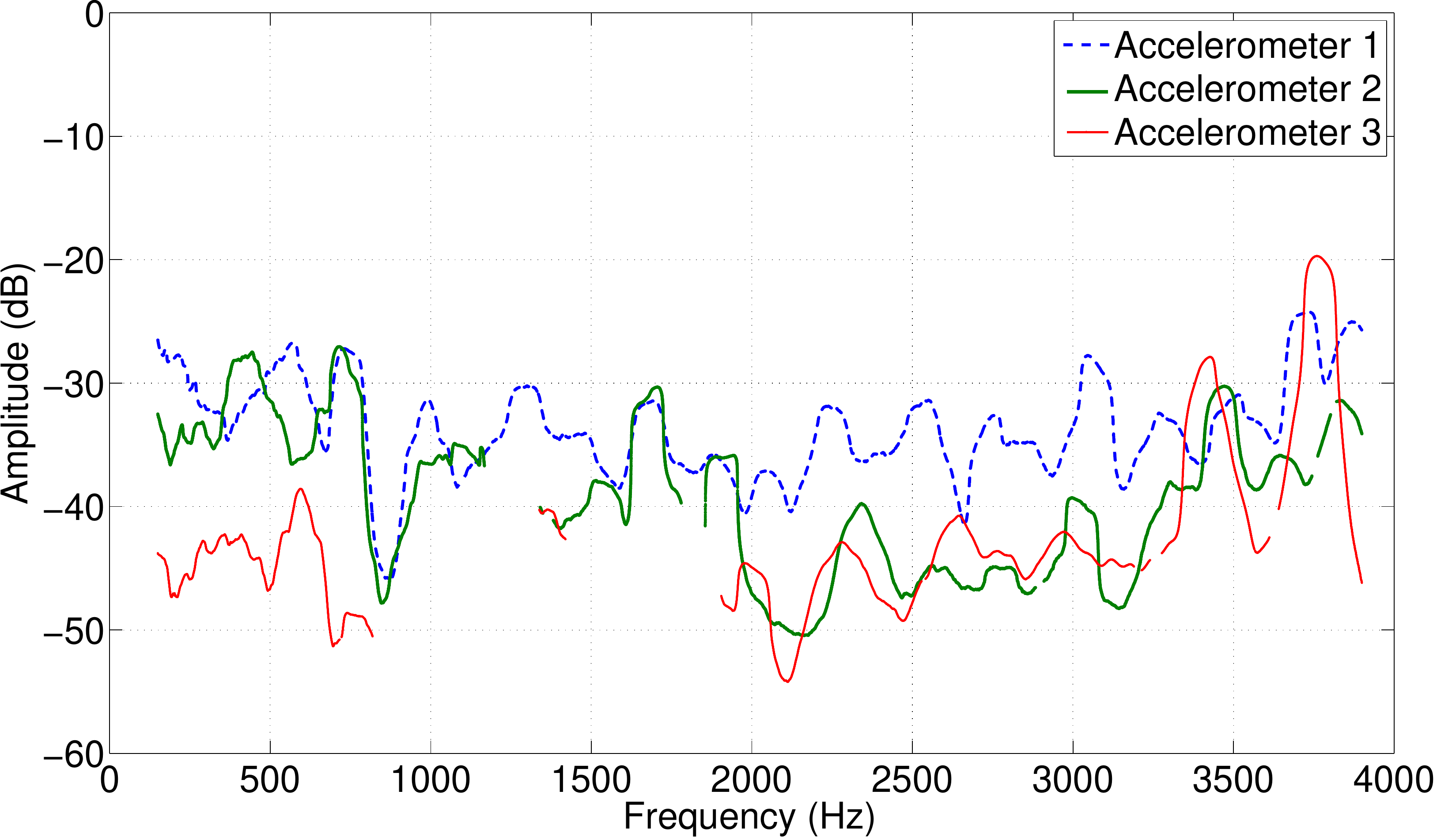}
  \caption{Estimated nonlinear contributions of the piano soundboard, computed according to Eq.~\eqref{eq:apport_NL_estG} and evaluated at three positions of the soundboard. {\color[rgb]{0,0,1}\hdashrule[0.5ex]{3em}{1pt}{2.5mm 1mm}}
: 
$C_{\mathcal{S}_G}(f)$ for A$_1$. {\color[rgb]{0,0.5,0}\hdashrule[.5ex]{3em}{1pt}{}}
~: 
$C_{\mathcal{S}_G}(f)$ for A$_2$. {\color[rgb]{1,0,0}\hdashrule[.5ex]{3em}{1pt}{}}
~: 
$C_{\mathcal{S}_G}(f)$ for A$_3$. An averaging window is applied to the amplitude of each spectrum so that each point corresponds to the average on a 100 Hz-bandwidth. See text for missing parts.}
\label{fig:NonLinPiano}
\end{center}
\end{figure}

The distortion rates of the piano soundboard $C_{\mathcal{S}_G}(f)$ at points A$_1$, A$_2$ and A$_5$ are shown in Fig.~\ref{fig:NonLinPiano}. The missing parts correspond to the anomalous situation described above. The apparent increase in nonlinearity near 4 kHz is probably an artefact of the method since the quality of the reconstruction of the nonlinear impulse responses is bad near the lower and upper bounds of the explored frequency range (50-4000 Hz in the present case). The three curves differ by not more than an order of magnitude, which is consistent with the fact that the level of the soundboard motion differs, but not widely, from place to place and therefore, that the responses in different locations do not carry the same amount of nonlinearity. Conversely, the fact that the three curves are roughly similar justifies, at least for the purpose of finding an order-of-magnitude, the approximations made above.

Altogether, the nonlinear part of the piano response appears to be contained within -30 to -50 dB. The order of magnitude of -40 dB can be retained for the total distortion rate at the \emph{ff} playing level.

%% file: ESPRIT_V42.tex
\section{Experimental modal analysis method}
\label{sec:Method}
The experimental study presented in this section aims at estimating the modal parameters (modal frequencies, modal dampings and modal shapes) of the upright piano soundboard in a wider than usual frequency range (see below). The terminology and symbols are the followings: when the soundboard vibrates freely, each mode corresponds to a time-signal of the generic form $\ee^{-\alpha t}\cos(2\pi\,f\,t+\varphi)$ where $f$ is the modal frequency and $\alpha$ is the modal damping, or damping rate. The modal loss-factor is defined by $\eta = \dfrac{\alpha}{\pi\,f}$ (also: twice the damping ratio and the reciprocal of the quality factor). The modal overlap $\mu$ is defined as the ratio between the half-power modal bandwidth and the average modal spacing: $\mu=\dfrac{\Delta f\idr{-3dB}}{\Delta f\idr{mode}}=\dfrac{f\,\eta}{\Delta f\idr{mode}}=\dfrac{\alpha}{\pi\,\Delta f}$.

The piano was put in a pseudo-anechoic room and excited either mechanically and impulsively with an impact hammer or continuously and acoustically with a loudspeaker. The former excitation yields the modal shapes but does not provide enough energy beyond a certain frequency, depending on the experimental conditions (here: \mbox{$\approx 500$ Hz}), hence the acoustical excitation (which does not give access to the modal shapes).

When a loudspeaker was used to excite acoustically the piano, the driving signal was adjusted in order to obtain a vibration level similar to the one used in the nonlinearity study; the linear contribution of the response was extracted prior to modal analysis. The impulsive excitation does not permit to separate the linear and nonlinear contributions. Nevertheless, in the light of the results presented above and considering the small amplitudes of displacement caused by the impacts on the soundboard (typically less than $8\cdot10^{-6}~\text{m}$, mostly due to a very low-frequency displacement and still less than 1/100 of the board thickness), we considered that the linear approximation was verified up to the precision of our measurements. In a way, this was confirmed by the fact that, in modal identifications, we never observe a frequency or damping rate that was exactly twice or three times that of a lower mode.

\begin{figure}[ht!]
\begin{center}
\includegraphics[width=1\linewidth]{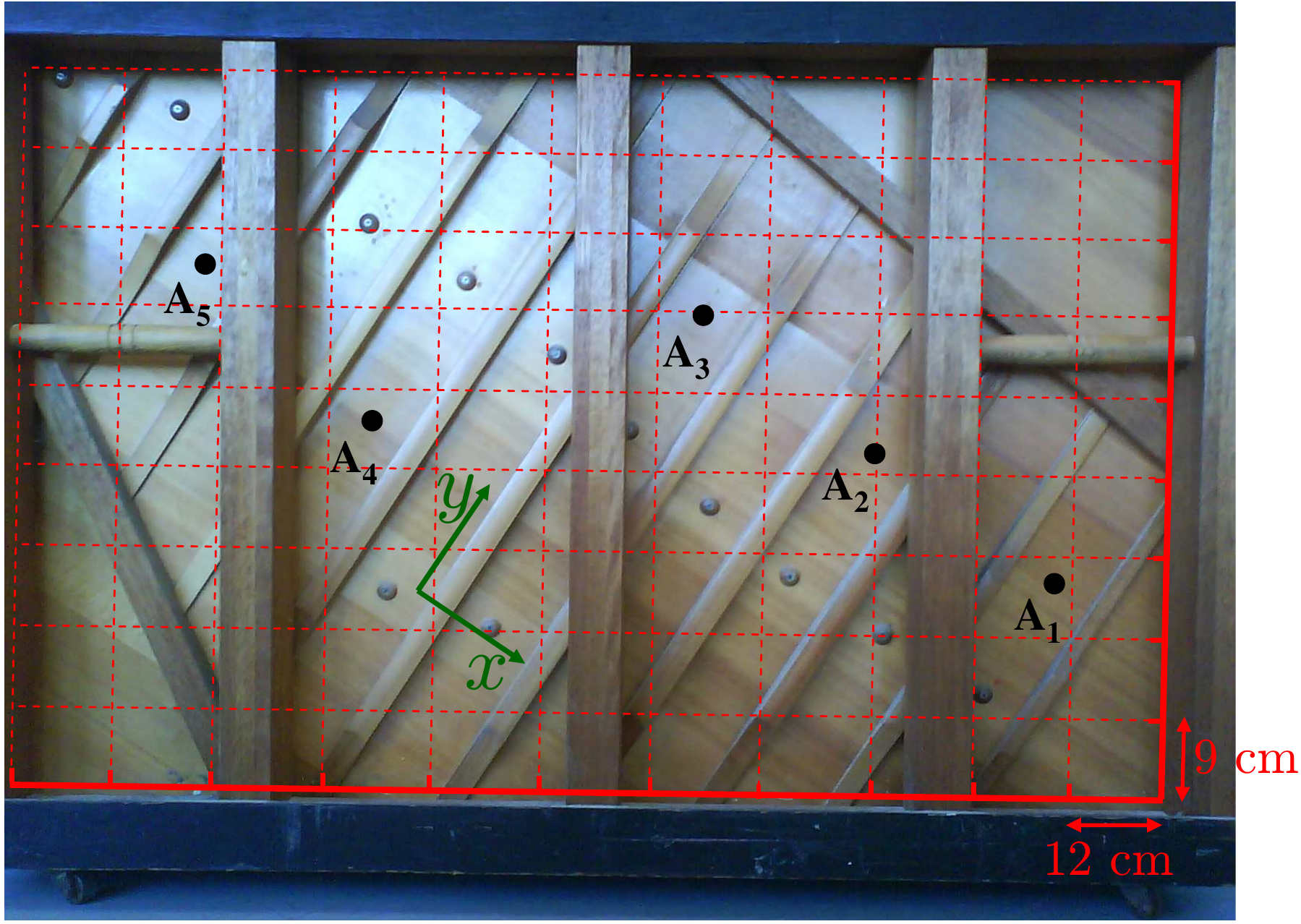}
\end{center}
\caption[aaa]{Rear view of the upright piano, with the mesh for modal analysis (in red) and the locations of the five accelerometers (in black).}
\label{fig:table_exp_maillage}
\end{figure}

The impact hammer (Kistler -- type 9722A) was struck at the nodes of a rectangular mesh of $12\times10$ points regularly spaced (Fig.~\ref{fig:table_exp_maillage}). The motion of the soundboard was measured with accelerometers ({two B\&K~2250A-10 and three B\&K~4393}) at five points in different zones of the board (Fig.~\ref{fig:table_exp_maillage}).

\begin{figure}[ht!]
\begin{center}
\includegraphics[width=1\linewidth]{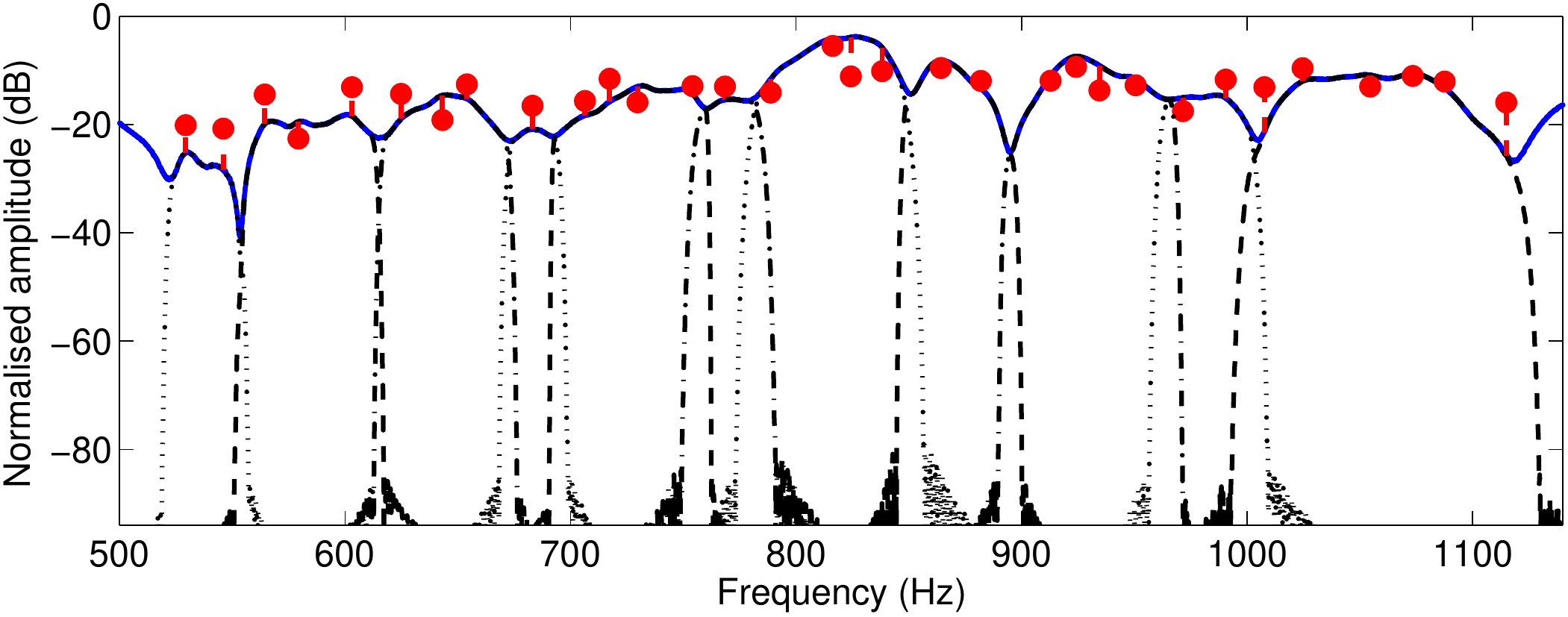}
\end{center}
\caption[aaa]{Typical bank-filtering analysis of a reconstructed impulse response between 550 and 1150~Hz (acoustical excitation). --~--~--\enskip,\enskip$\cdots$\enskip,\enskip--\,$\cdot$\,--\enskip:~amplitude responses of the (slightly overlapping) narrow-band filters. {\color[rgb]{0,0,1}------}\,: Fourier spectrum of the impulse response at point \textbf{A}$_\mathbf{2}$. {\color[rgb]{1,0,0}$\bullet$\,}: modes estimated by the high-resolution modal analysis (modal amplitudes and frequencies).}
\label{fig:spectre_500_1200_compoOK}
\end{figure}

%
For the acoustical excitation of the soundboard, the procedure is the same as in Sec.~\ref{sec:Linearity} (an exponential swept-sine [50-4000]~Hz with a 40 kHz sampling frequency, a $T=26~\text{s}$ duration and an assumed $N=4$ maximum order of nonlinearity). The impulse response of the soundboard is reconstructed by the deconvolution technique described in~\cite{EGE2009} and analysed through a series of band-pass filters (a typical bank filtering analysis is displayed in Fig.~\ref{fig:spectre_500_1200_compoOK} between 550 and 1150~Hz). The cut-off frequencies of the finite-impulse-response (FIR) filters were chosen at local minima of the Fourier spectrum of the response. If necessary, when there is a doubt on the number of components in one frequency-band, two successive overlapping filters were occasionally chosen.

In the piano, the modal overlap factor $\mu$ exceeds 30\% for all frequencies above 150 Hz. In other words, almost the whole frequency range of interest is outside the low-frequency range where the usual modal analysis technique, based on Fourier-analysis, is applicable. With such modal overlap factors, the estimation of a damping rates by the width of the peak in the spectrum is not possible either. The modal behaviour of the soundboard of the upright piano has been investigated by means of a recently published high-resolution modal analysis technique~\cite{EGE2009} which avoids the frequency-resolution limitations of the Fourier transform.

\begin{figure}[ht!]
\begin{center}
\includegraphics[width=1\linewidth]{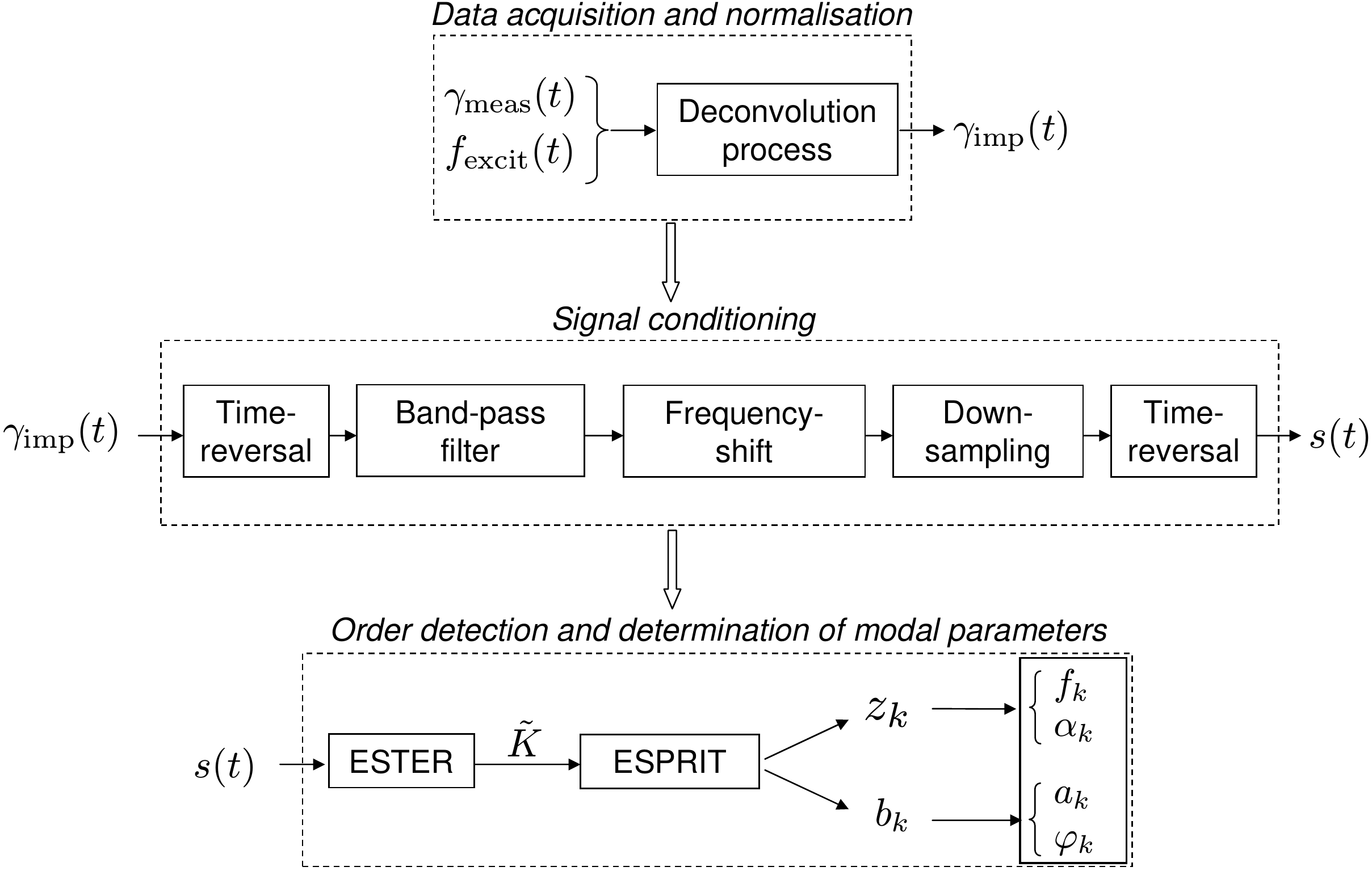}
\end{center}
\caption[blockdiagram]{Block diagram of the high-resolution modal analysis method with which the modal frequencies $f_k$, modal dampings $\alpha_k$, modal amplitudes $a_k$ and phases $\varphi_k$ are determined (after~\cite{EGE2009}).}
\label{fig:blockdiagram}
\end{figure}

The modal analysis technique used here is well suited for structures made of moderately damped materials such as spruce, and for frequencies where the modal overlap lies between 30\% and 100\%. Based on the ESPRIT algorithm~\cite{ROY1989}, it assumes that the signal is a sum of complex exponentials and white noise; it projects the signal onto two subspaces: the subspace spanned by the sinusoids (signal subspace) and its supplementary (noise subspace). The rotational invariance property of the signal subspace (see~\cite{ROY1989} for details) is used to estimate the modal parameters: frequencies, damping factors and complex amplitudes. The dimensions of both subspaces must be chosen~\emph{a priori} and the quality of the estimation depends significantly on a proper choice for these parameters. The best choice for the dimension of the modal subspace is the number of complex exponentials actually present in the signal. This number ($\tilde{K}$ in Fig.~\ref{fig:blockdiagram}) is twice the number of decaying sinusoids. Prior to the modal analysis itself, an estimate of this number is obtained by means of the ESTER technique~\cite{BAD2006} which consists in minimising the error on the rotational invariance property of the signal subspace spanned by the sinusoids. The block diagram given in Fig.~\ref{fig:blockdiagram} describes the three main steps of the modal analysis method: (a) reconstruction of the acceleration impulse response, (b) signal conditioning, (c) order detection, (d) determination of modal parameters. 

\renewcommand{\figureplace}{%
\begin{center}
[\figurename~\thepostfig\ about here (with ref.~\cite{EGE2009} in caption).]
\end{center}}

\renewcommand{\figureplace}{%
\begin{center}
[\figurename~\thepostfig\ about here.]
\end{center}}

\begin{figure}[ht!]
\begin{center}
\includegraphics[width=1\linewidth]{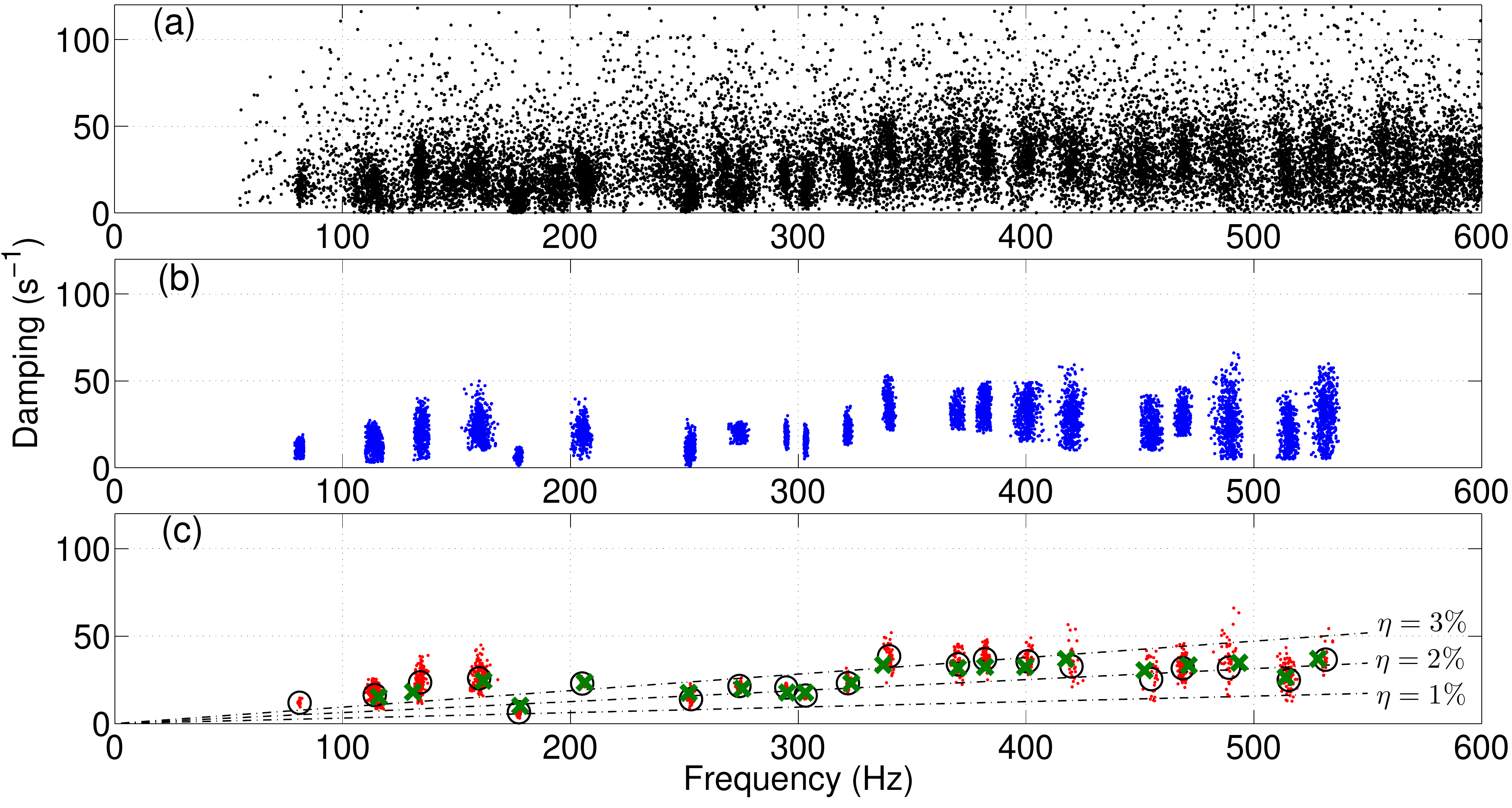}
\end{center}
\caption[aaa]{Modal frequencies and dampings in the \mbox{[0-600] Hz} frequency-band, obtained after an \emph{impulse} excitation and given by a high-resolution modal analysis. (a)~Direct analysis. (b)~Analysis including a narrow band-pass filtering. (c)~Results after suppression of the (usually poor) estimations in the nodal regions. ($\circ$)~: retained modal parameters. ({\color[rgb]{0,0.5,0}$\times$})~: weighted means of the modal parameters estimated at four points of the soundboard with an \emph{acoustical} excitation (see following section). --$\,\cdot\,$--~: constant loss-factors $\eta~$=\,1, 2 and 3\%.} 
\label{fig:res_esprit_piano}
\end{figure}

\begin{figure}[ht!]
\includegraphics[width=1\linewidth]{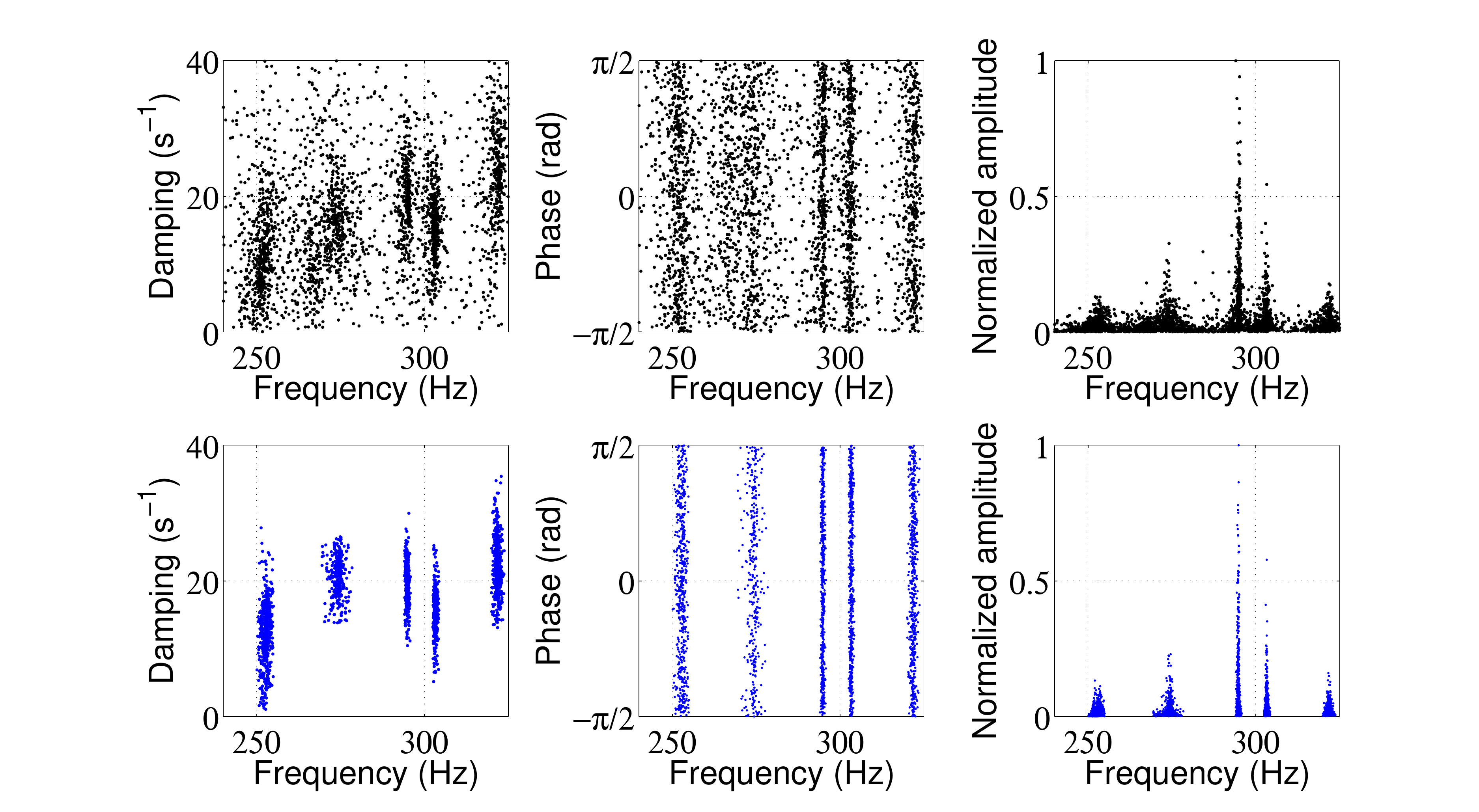}
\caption[aaa]{Necessity of the narrow band-pass filtering step. Results for five modes in the \mbox{[230-330] Hz} frequency-band before filtering (top diagrams), after filtering (bottom diagrams).}
\label{fig:esprit_piano_bandeetroite_2}
\end{figure}

Below 500 Hz, for each of the $120\times5$ measurements, reconstructing and analysing the impulse response as described in Fig.~\ref{fig:blockdiagram} yields the results presented in the top frames of Figs.~\ref{fig:res_esprit_piano} and \ref{fig:esprit_piano_bandeetroite_2}. In order to measure the damping with enough precision, it proved necessary to band-filter the impulse responses prior to analysis, yielding the results presented in the middle and bottom frames of Figs.~\ref{fig:res_esprit_piano} and \ref{fig:esprit_piano_bandeetroite_2} respectively. Finally, the modal dampings were extracted by averaging the results after suppression of the (usually poor) estimations in the nodal regions: bottom frame of Fig.~\ref{fig:res_esprit_piano}.

Above 550~Hz, the clusters in Fig.~\ref{fig:res_esprit_piano}(a) are more difficult to identify, owing to a too low Signal-to-Noise Ratio (SNR~$\approx35$~dB): the excitation by an impulse hammer is intrinsically limited in frequency since the force exerted by the hammer is of finite duration (in the order of the time taken by the initial impulse to be echoed by the closest discontinuity). Technically, the signal-to-noise ratio is not high enough beyond \mbox{$\approx 500$ Hz} for determining modal parameters with enough precision. In order to extend the estimation of the modal frequencies and dampings above 500~Hz, we replaced the impulsive mechanical excitation by a continuous acoustical one (Fig.~\ref{fig:piano_chambre}). The impulse responses were extracted and the modal frequencies and dampings were estimated by the procedure presented in Fig.~\ref{fig:blockdiagram} but the modal shapes could not be determined.

The results are presented in Sec.~\ref{sec:Results} and discussed together with the results of the numerical modal analysis.

%% file: FEM_V42.tex
\section{Finite-element model of the piano soundboard}
\label{sec:FEM}

\begin{figure}[!ht]
\begin{center}
\includegraphics[angle=-90,width=0.8\linewidth]{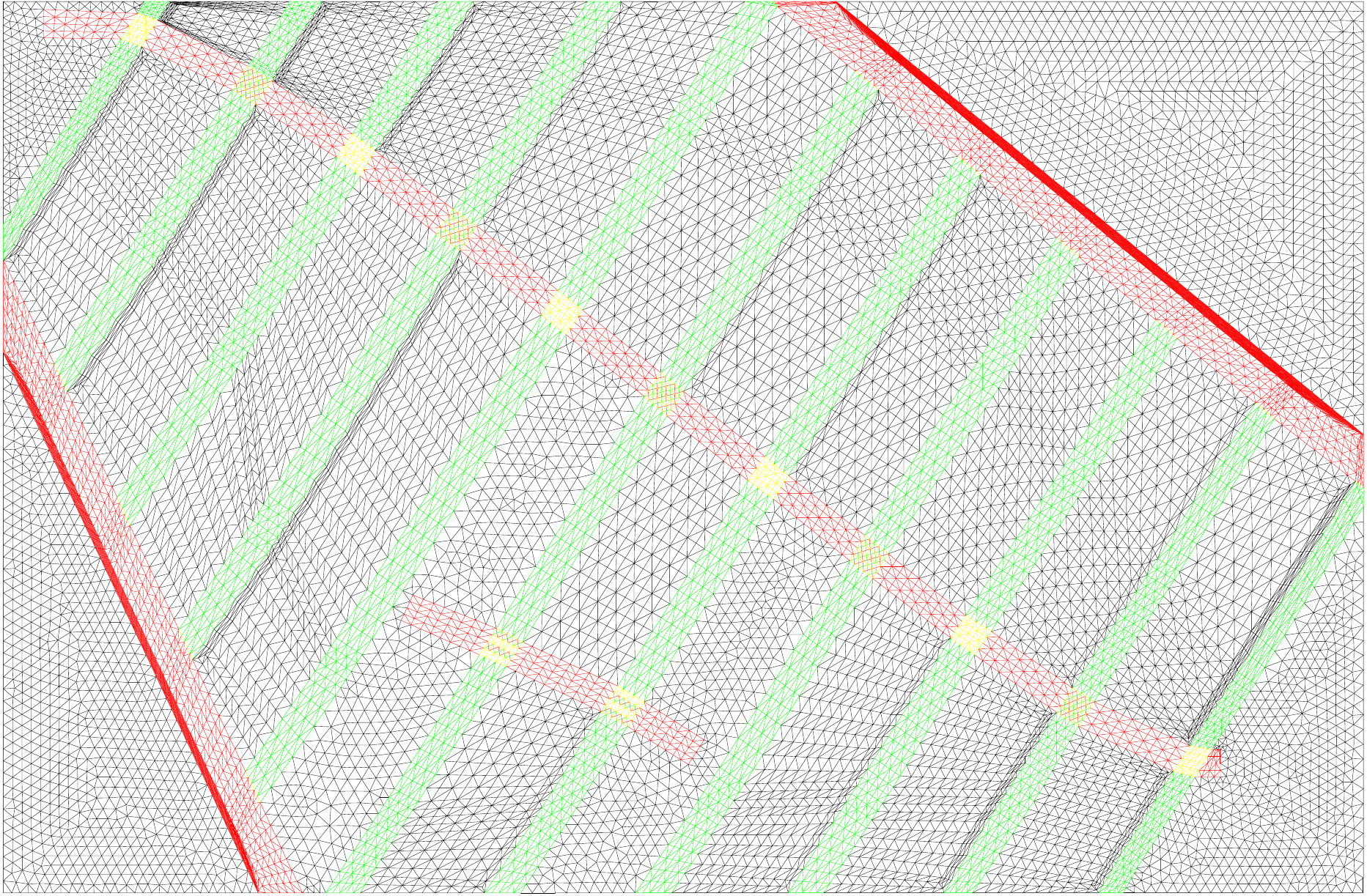}
\end{center}
\caption[Mesh]{Mesh of the soundboard: wood panel (black), bars (red), ribs (green) and bridges (red). The overall dimensions of the rectangle are \mbox{0.91 m $\times$ 1.39~m}. The thickness of the wood panel (without ribs) is 8~mm. The angle between the long-side of the soundboard and the "L"-direction of the panel wood (along the grain) is 33$^\text{o}$.}
\label{fig:Mesh}
\end{figure}

A finite-element model (FEM) of the soundboard has been written by means of the free software Cast3M~\cite{CASTEM}. The soundboard model is that of a rectangular 
plate with two strong bars delimiting two so-called "cut-off corners" in opposite angles, 11 thin bars (ribs) and the two bridges. Following makers habit, the structure is given the shape of a spherical shell, here of radius $R=43$~m, corresponding to a value of \mbox{$\approx$ 8 mm} for the so-called \emph{crown} at the centre of the soundboard. This value of the crown is consistent with Conklin's observations~\cite{CON1996_2}, corresponds approximately to the plate thickness and represents a standard value according to discussions with piano manufacturers. Different manufacturing processes can be employed to realize crowning, all leaving residual stresses at the time when the soundboard is fixed in its rim, before applying string loading. In the absence of documentation, we have chosen to ignore residual stresses here.

Due to the tension of the strings and to the angle that they form with the plane of the soundboard when going through the bridge, a load (called \emph{downbearing}) is exerted on the bridge, in the direction perpendicular to the soundboard. This load results in an internal stress and has been included in the numerical model as a vertical (transverse) force uniformly distributed along the two bridges. This load can be adjusted to some extent by the maker and has been chosen here so that the crown at the centre of the soundboard is reduced to one-half of its initial (without string loading) value: this is also a standard reduction based on piano manufacturers know-how.

The geometrical data were measured directly on the soundboard (see the caption of Fig.~\ref{fig:Mesh}). All finite elements are triangular thin-shell orthotropic elements. The mesh (Fig.~\ref{fig:Mesh}) has 14267 nodes. The shell thickness is 8~mm; the height of the medium-bridge is 3~cm and that of the bass-bridge is 5.5~cm. The cut-off bars are 3.3~cm thick. The dimensions of the ribs and the inter-rib distances are given in  Appendix~\ref{sec:RibDimensions}. The boundaries of the soundboard are supposed to be clamped and the dynamics is supposed to be conservative.

All the pieces are made of wood, considered here as an orthotropic material: spruce for the rectangular panel and the ribs, fir for the cut-off bars, maple for the bridges. For each of these three wood species, four elastic constants are necessary to model the chosen finite elements. Pianos are not all made of woods with exactly the same characteristics. We have retained the maple characteristics given by Haines~\cite{HAI1979} and the fir characteristics given by Berthaut~\cite{BER2004}. Although a parametric study falls beyond the scope of this article, we have run the FEM simulations with three sets of values for spruce characteristics as given by Haines~\cite{HAI1979} (Norway spruce) and Berthaut~\cite{BER2004} (Sitka spruce). The French piano maker Stephen Paulello gave us information on spruce characteristics that he had observed on ordinary pianos: we refer to it as "mediocre spruce". Since the  piano that has been chosen here is clearly not a high-end model of a well-known brand, the latter wood is plausible; in addition, the numerical results also fit best the experimental results, in terms of modal density (see Fig.~\ref{fig:densitemodale_guidedonde}). The corresponding numerical values and that of $\rho$ are given in Table~\ref{tab:caracmeca_num}. The "L" (longitudinal) direction refers to that of the grain and corresponds to the main axis of orthotropy with the higher elasticity modulus. The "R" (radial) direction is across the grain and corresponds to the other main axis of orthotropy. The ribs are cut in the "L" direction of their wood. They are glued on the panel in the "R" direction of the panel's spruce, corresponding to $O_y$. The cut-off bars and the two bridges are also cut in their "L" direction.

In order to obtain the same crown under string loading for each set of spruce characteristics, and owing to the possible adjustment of the downbearing by the piano makers, the overall static force value was set to \mbox{2200 N}, \mbox{1700 N} and \mbox{1400 N} for the Norway spruce, the Sitka spruce and the mediocre spruce respectively.

\renewcommand{\tableplace}{%
\begin{center}
[\tablename ~\theposttbl\ about here (with refs.~\cite{BER2004,HAI1979} in caption).]
\end{center}}

\begin{table}[ht!]
\begin{center}
\begin{tabular}{|c||c|c|c|c|c|c|c|c|c|c|c|}
\hline
\  & $E_\text{L}$ (GPa)& $E_\text{R}$ (GPa)& $G_\text{LR}$ (GPa)& $\nu_\text{LR}$ & $\rho$~(kg~m$^{-3}$)\\
\hline

\hline
Fir & 8.86 & 0.54 & 1.60 & 0.3 & 691\\
\hline
Maple & 10.0 & 2.20 & 2.0 & 0.3 & 660\\
\hline
Norway spruce & 15.80 & 0.85 & 0.84 & 0.3 & 440\\
\hline
Sitka spruce & 11.50 & 0.47 & 0.5 & 0.3 & 392\\
\hline
Mediocre spruce & 8.80 & 0.35 & 0.5 & 0.3 & 400\\
\hline
\end{tabular}
\end{center} 
\caption[aaa]{Density and elastic constants of fir (cut-off bars), maple (bridges) and spruce (ribs and panel). The data of the first and fourth lines are given by Berthaut~\cite{BER2004}, those of the second and third lines by Haines~\cite{HAI1979}, that of the last line by Paulello. The subscripts $_\text{L}$ and $_\text{R}$ stand for "longitudinal" and "radial" respectively. The radial and longitudinal directions refer to how strips of wood are cut and correspond to the "along the grain" and "across the grain" directions respectively. In the geometry of the soundboard (Figs.~\ref{fig:board_twofaces}, \ref{fig:table_exp_maillage} and \ref{fig:Mesh}), the $x$- and $y$- directions correspond to $_\text{L}$ and $_\text{R}$ respectively for the spruce panel: $E\idr{L}=E_x,\ E\idr{R}=E_y$.
}
\label{tab:caracmeca_num}
\end{table}

\renewcommand{\tableplace}{%
\begin{center}
[\tablename ~\theposttbl\ about here.]
\end{center}}

%% file: Results_V43.tex
\section{Results and discussion}
\label{sec:Results}
Results are presented and discussed in terms of modal dampings (Sec.~\ref{sec:dampings}), first modal shapes, boundary conditions and first modal frequencies (Sec.~\ref{sec:shapes}). Finally, the modal density $n(f)$, defined as the number of modes per Hz, or reciprocal of the average frequency-interval between two modes is discussed in Sec.~\ref{sec:density}.

\subsection{Modal dampings}
\label{sec:dampings}
\begin{figure}[ht!]
\begin{center}
\includegraphics[width=1\linewidth]{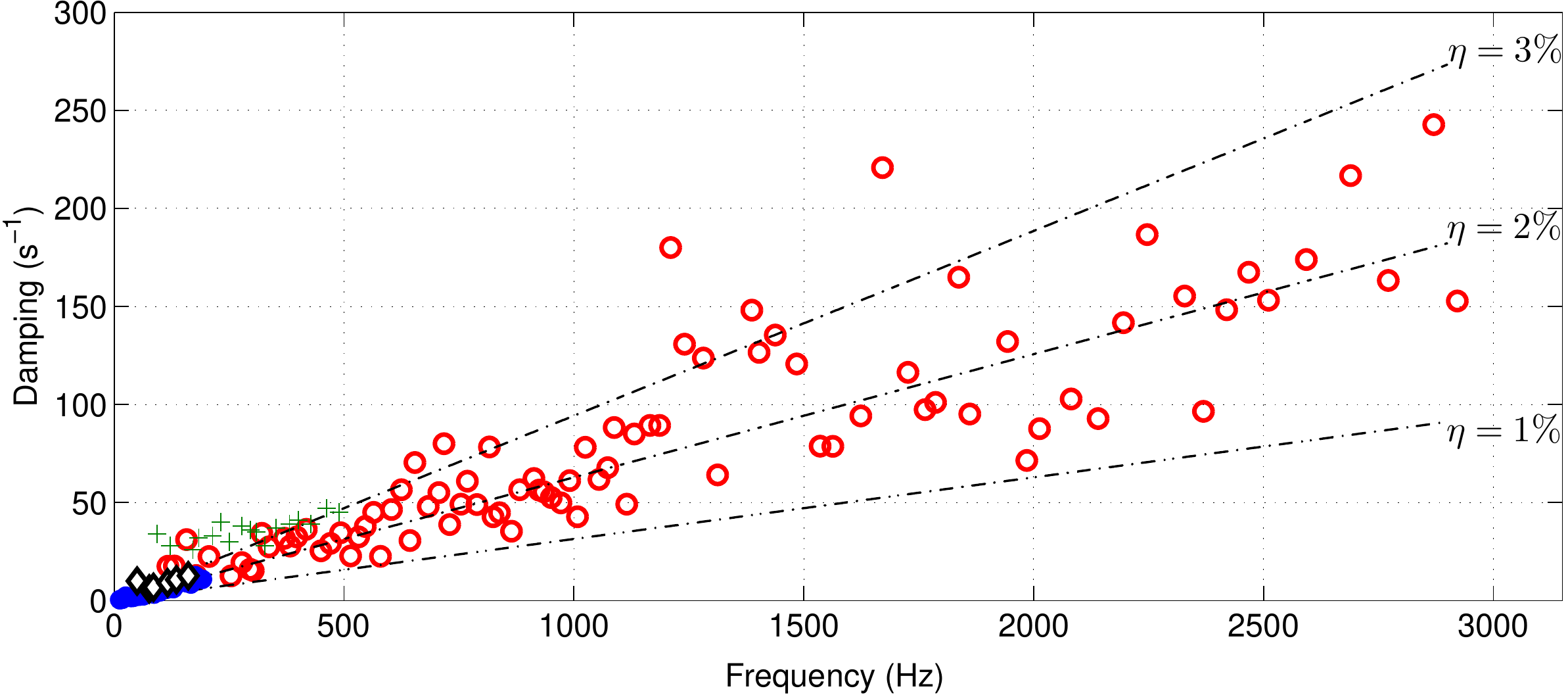}
\end{center}
\caption[aaa]{Modal frequencies (abscissa) and damping factors (ordinates). ({\color[rgb]{1,0,0}{$\circ$}})~: estimations at point~\textbf{A}$_\mathbf{2}$ of the soundboard. Bibliographical results: {{$\diamond$}}~\cite{SUZ1986}; ({\color[rgb]{0,0.5,0}{$+$}})~\cite{DER1997}; ({\color[rgb]{0,0,1}{$\bullet$}})~\cite{BER2003}. --$\,\cdot\,$--~: constant loss-factor curves $\eta~$=\,1, 2 and 3\%.}
\label{fig:res_esprit_amort_500_30008_compar}
\end{figure}

The modal dampings are reported up to 500 Hz in the bottom frame of Fig.~\ref{fig:res_esprit_piano} and up to 3~kHz in Fig.~\ref{fig:res_esprit_amort_500_30008_compar}, together with bibliographical results. The observed values yield values of the modal overlap $\mu$ ranging from around 30\% at 150~Hz to around 70\% at 550~Hz: this explains why the bibliographical results on modal dampings, obtained by modal analyses based on the Fourier transform, are limited to \mbox{$\approx$ 500 Hz}. The frequency-domain explored here includes mid-frequencies, which makes the acoustical excitation technique combined with the high-resolution analysis very appealing for modal analyses of musical instruments.

Except for the first four low-frequency resonances (see Fig.~\ref{fig:res_esprit_piano}-c), at which the energy losses at the rim are probably not negligible compared to those inside wood, the modal loss-factors up to around 1200~Hz range from 1\% to 3\% (mean of \mbox{$\eta\approx2.3\%$} for the 55 lowest-frequency estimations). This corresponds roughly to what they would be if losses were located in spruce only, where loss factors lie commonly between 1 and 3\%.

\renewcommand{\figureplace}{%
\begin{center}
[\figurename~\thepostfig\ about here (with refs.~\cite{SUZ1986,DER1997,BER2003} in caption).]
\end{center}}

\renewcommand{\figureplace}{%
\begin{center}
[\figurename~\thepostfig\ about here.]
\end{center}}

At higher frequencies, modal dampings increase from a mean value of about $80~\text{s}^{-1}$ below 1200~Hz to about $130~\text{s}^{-1}$ between 1200 and 1500~Hz. This is probably due to a change in the proportion of the energy lost in wood to the acoustically radiated energy. Interestingly, Suzuki~\cite{SUZ1986} noticed on a small grand that "\emph{the transition range from less efficient to efficient sound radiation is 1-1.6~kHz}". Above 1.8~kHz (or slightly less than that), the loss factors are again in the order of the material loss-factors for spruce.

\subsection{Modal shapes, boundary conditions and modal frequencies}
\label{sec:shapes}
\begin{figure}[!ht]
\begin{center}
\scriptsize
\begin{tabular}{cccc}
\hspace{0.24\linewidth}&%
\includegraphics[width=0.24\linewidth]{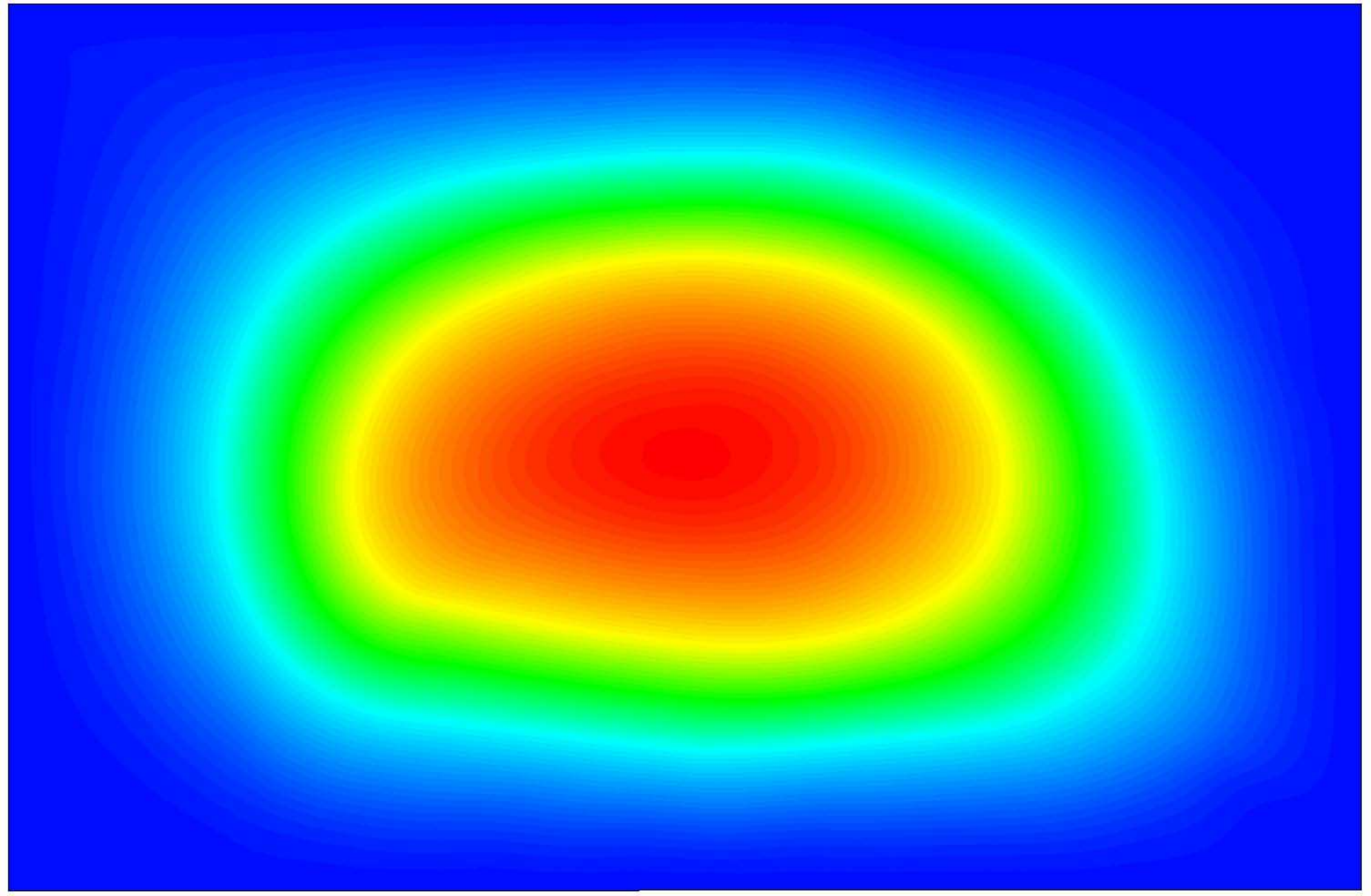}&%
\multicolumn{2}{c}{\includegraphics[width=0.24\linewidth]{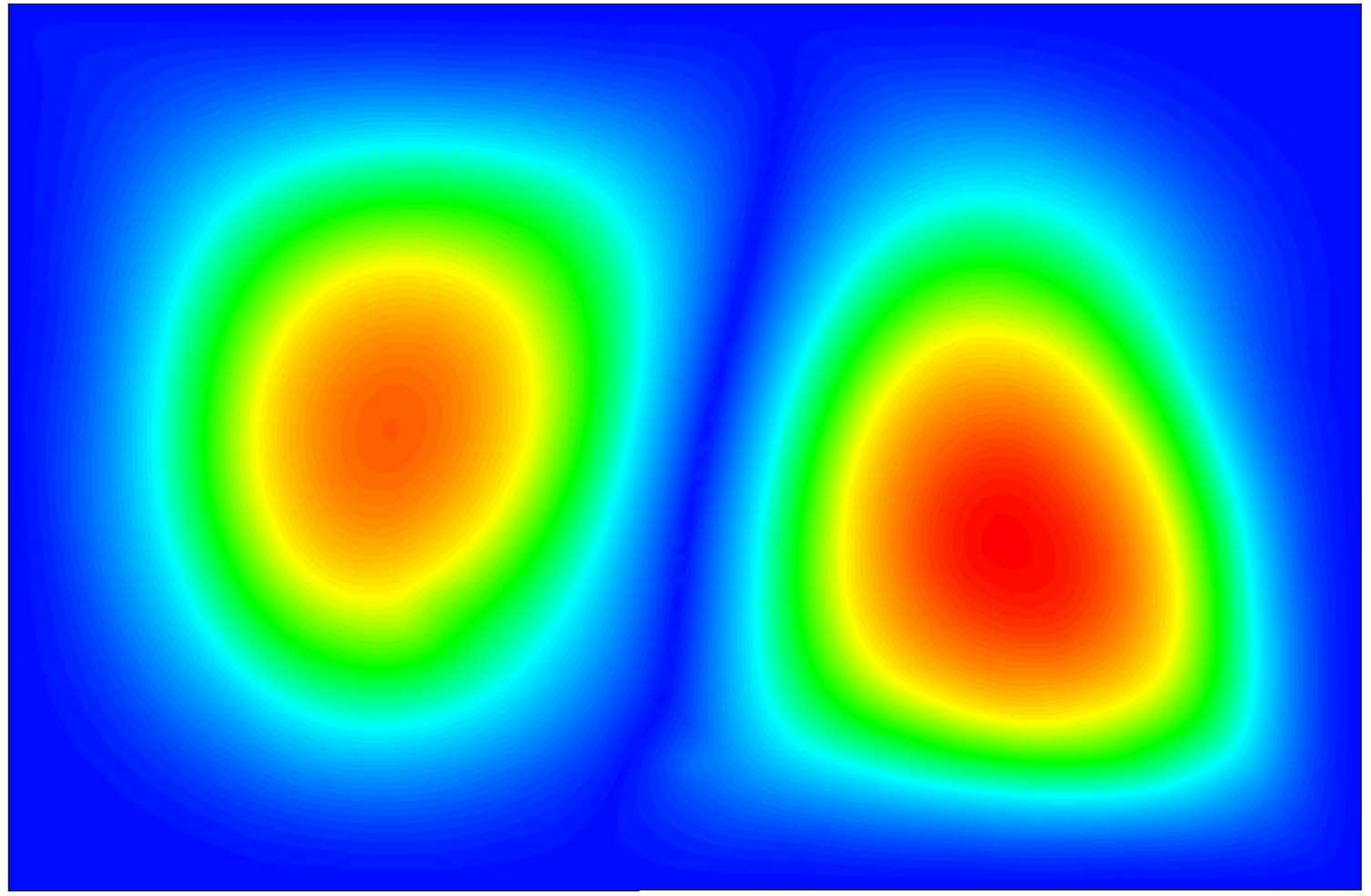}}%
\vspace*{-.10cm}\\
\vspace*{-.15cm}%
&84 Hz (Ns), 78 Hz (Ss), 71 Hz (ms)&\multicolumn{2}{c}{127 Hz (Ns), 119 Hz (Ss), 109 Hz (ms)}\\
\vspace*{-.15cm}%
Mode MMCM &Mode (1,1) &\multicolumn{2}{c}{Mode (2,1)}\\
81 Hz &114 Hz &\multicolumn{2}{c}{$\overbrace{\text{134 Hz\hspace{0.24\linewidth}159 Hz}}$}%
\vspace*{0.05cm}\\
\includegraphics[width=0.24\linewidth]{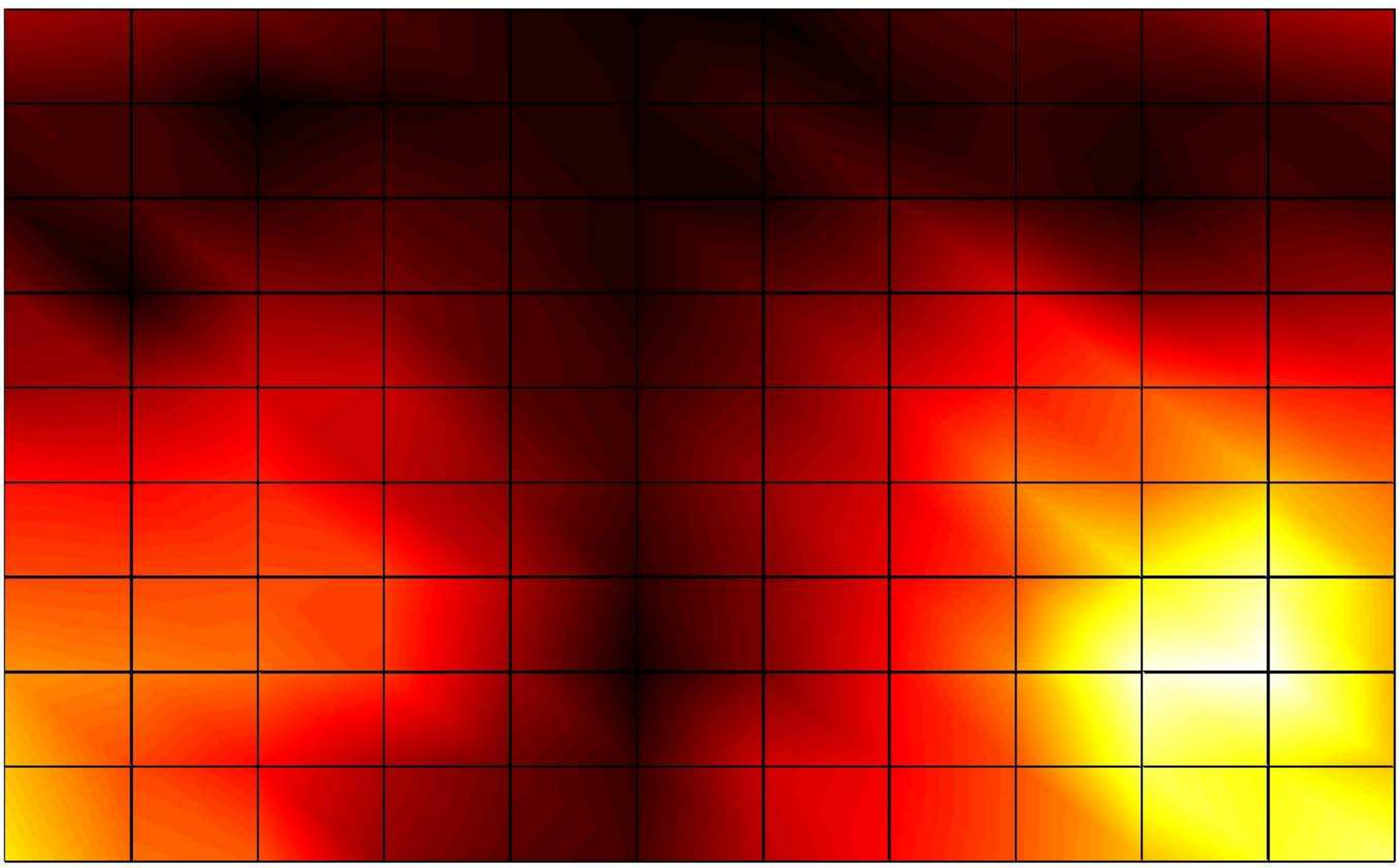}&%
\includegraphics[width=0.24\linewidth]{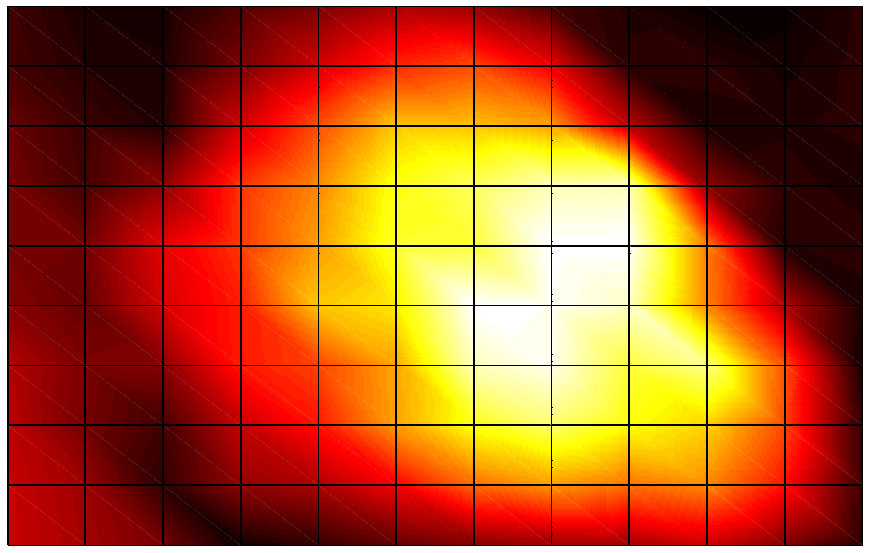}&%
\includegraphics[width=0.24\linewidth]{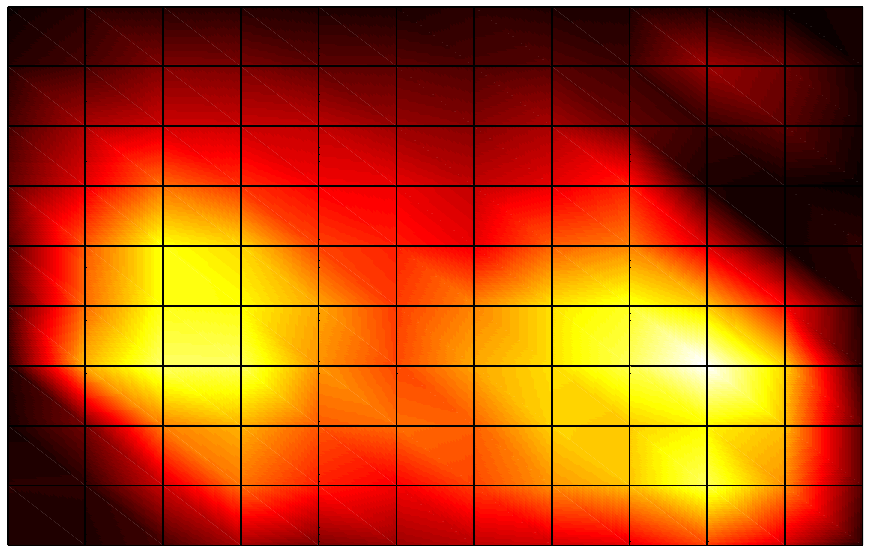}&%
\includegraphics[width=0.24\linewidth]{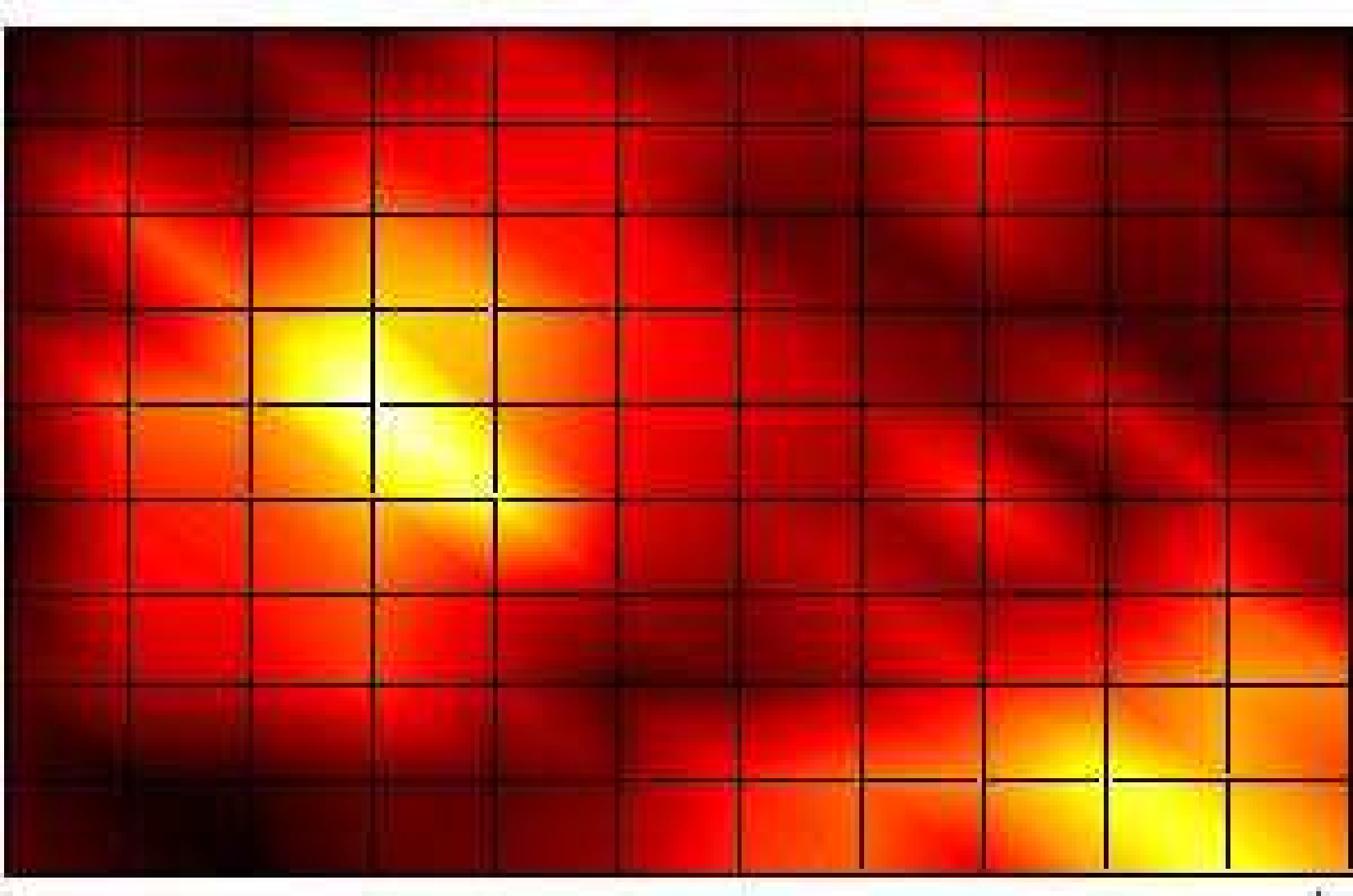}%
\vspace*{.3cm}\\%
\hline
\end{tabular}

\vspace*{.3cm}
\begin{tabular}{cccc}
\multicolumn{2}{c}{\includegraphics[width=0.24\linewidth]{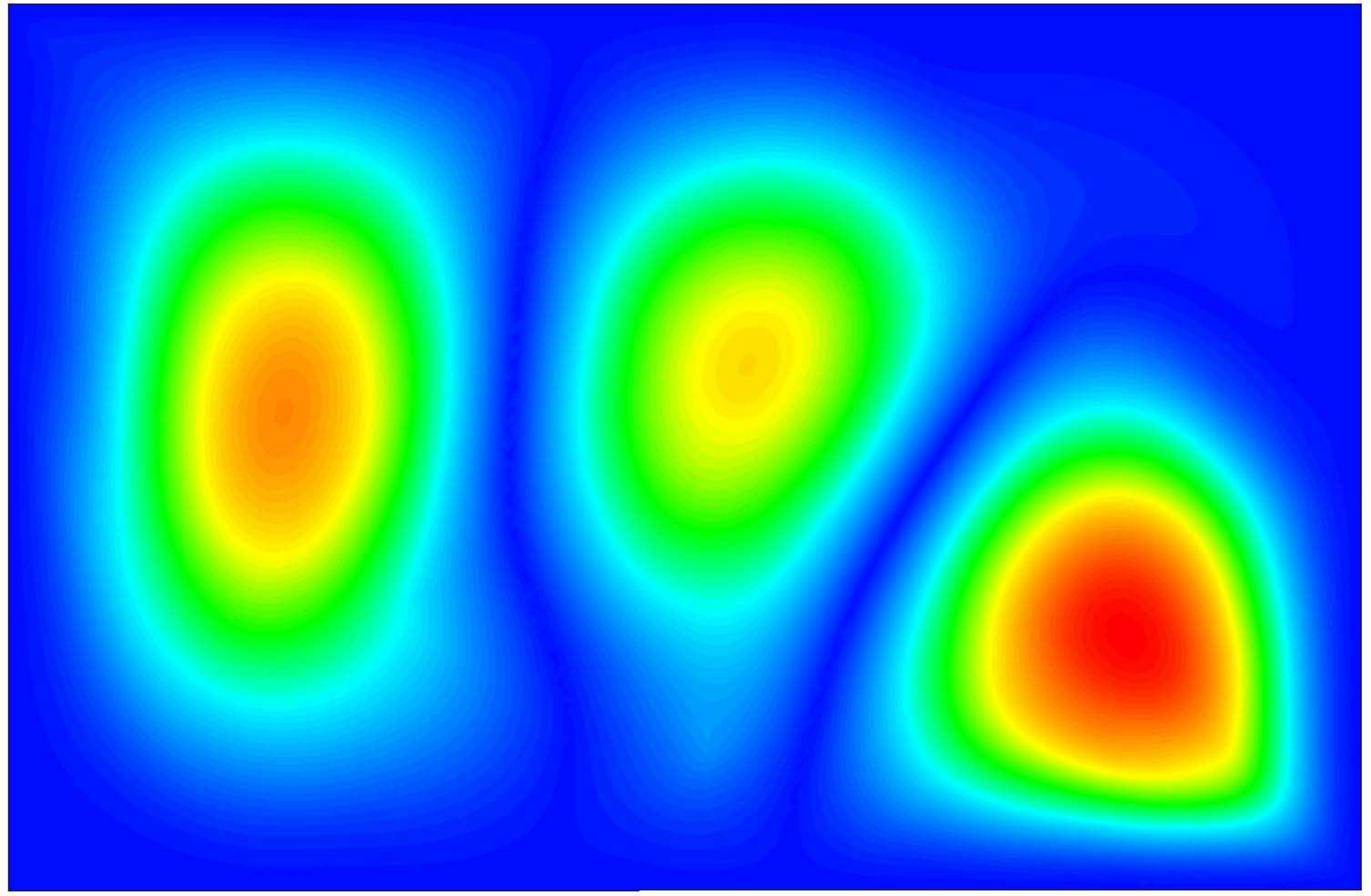}}&%
\includegraphics[width=0.24\linewidth]{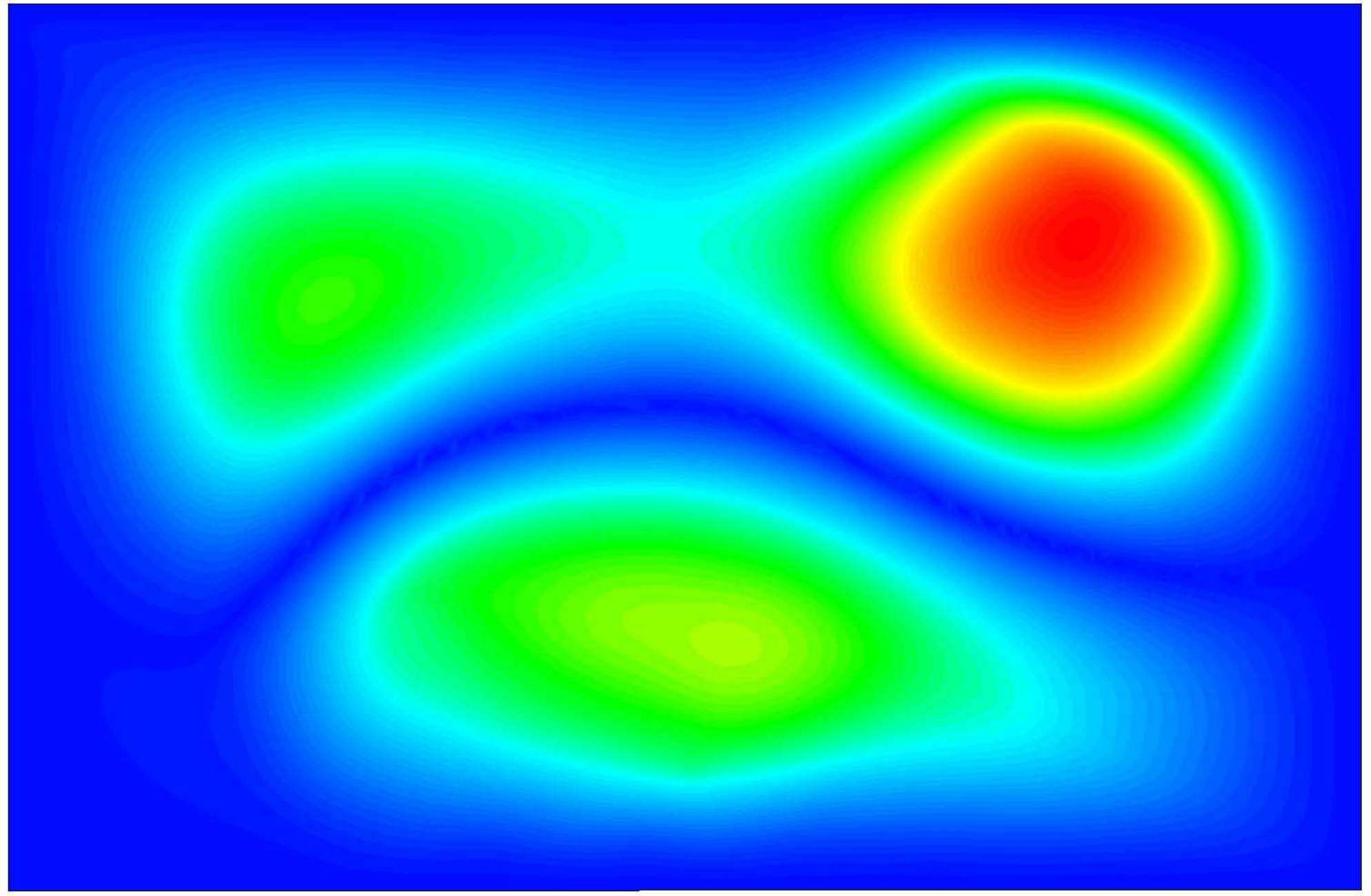}&%
\includegraphics[width=0.24\linewidth]{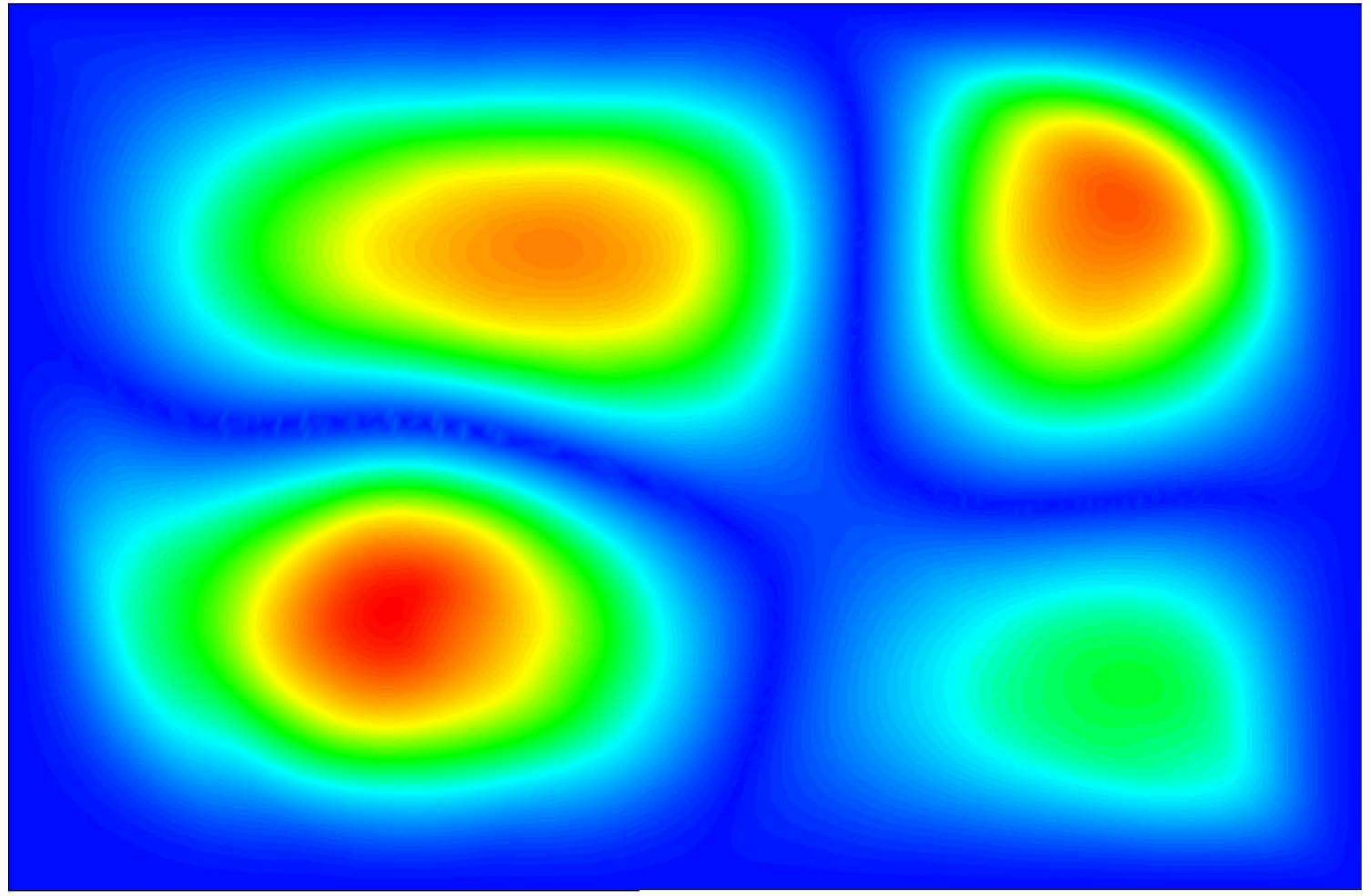}%
\vspace*{-.10cm}\\
\multicolumn{2}{c}{\vspace*{-.15cm}%
194 Hz (Ns), 182 Hz (Ss), 168 Hz (ms)} &204 Hz, 192 Hz, 176 Hz &229 Hz, 213 Hz, 199 Hz \\%
\multicolumn{2}{c}{\vspace*{-.15cm}Mode (3,1)} &Mode (1,2) & Mode (2,2)\\%
\multicolumn{2}{c}{$\overbrace{\text{177 Hz\hspace{0.24\linewidth}205 Hz}}$} &253 Hz &%
\vspace*{0.05cm}\\
\includegraphics[width=0.24\linewidth]{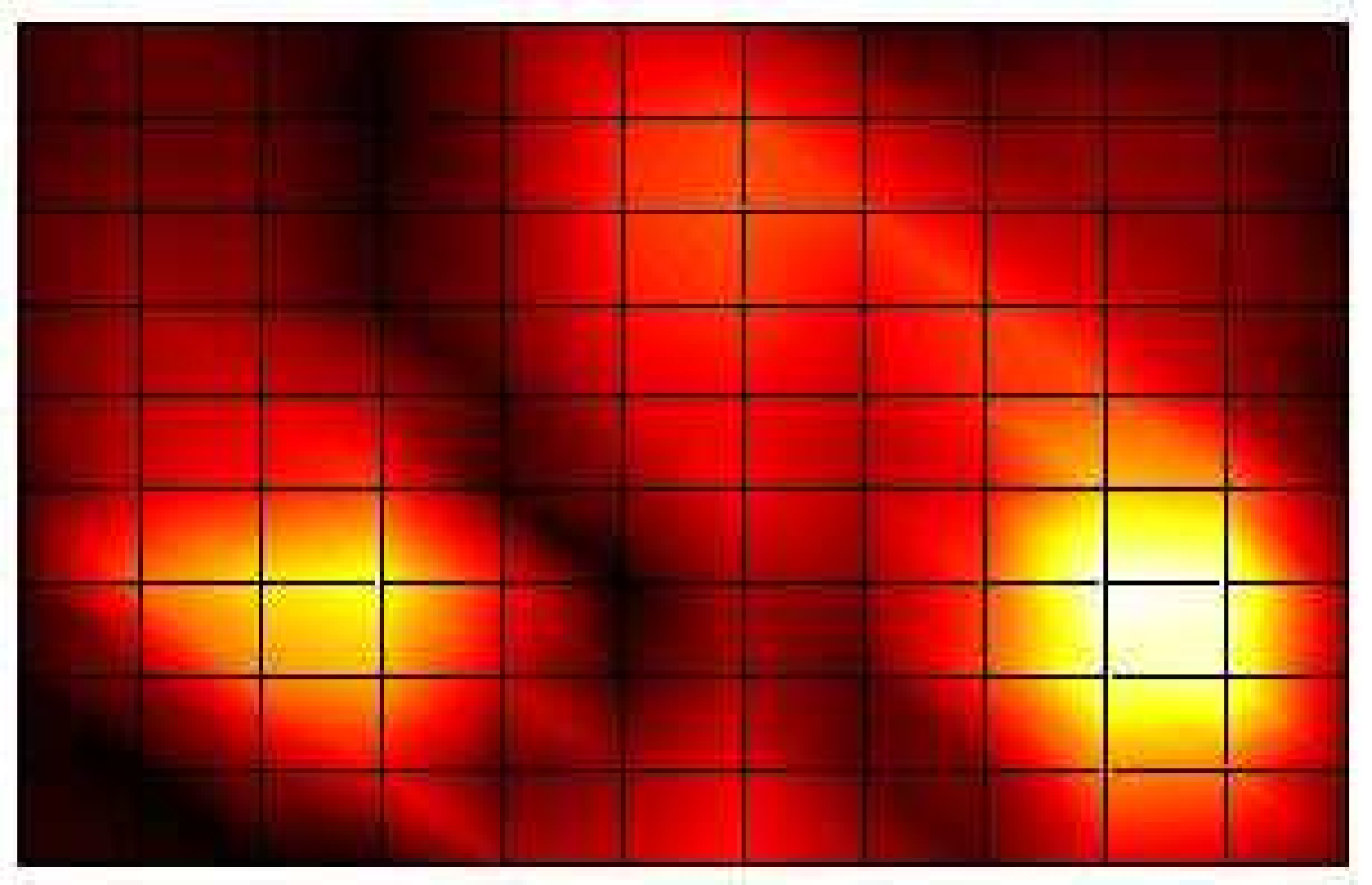}&%
\includegraphics[width=0.24\linewidth]{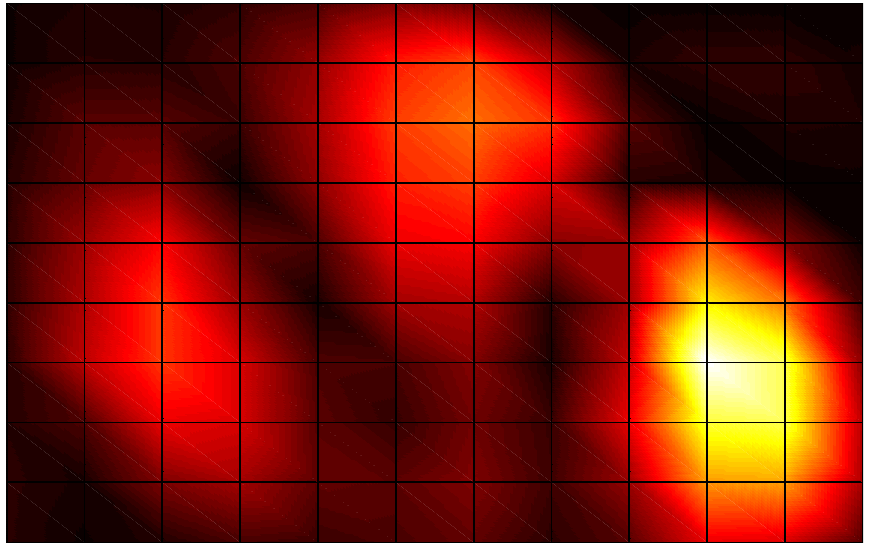}&%
\includegraphics[width=0.24\linewidth]{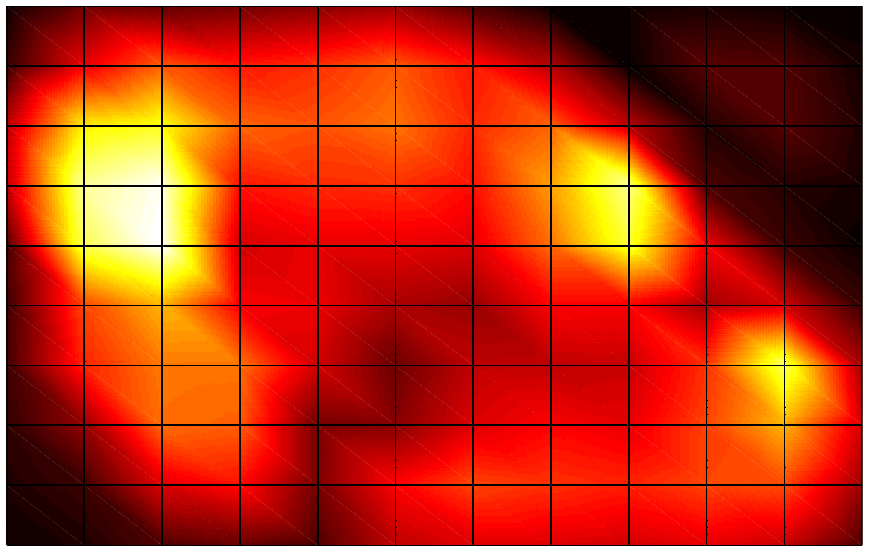}&%
\vspace*{.3cm}\\%
\hline
\end{tabular}

\vspace*{.3cm}
\begin{tabular}{ccc}
\multicolumn{2}{c}{\includegraphics[width=0.24\linewidth]{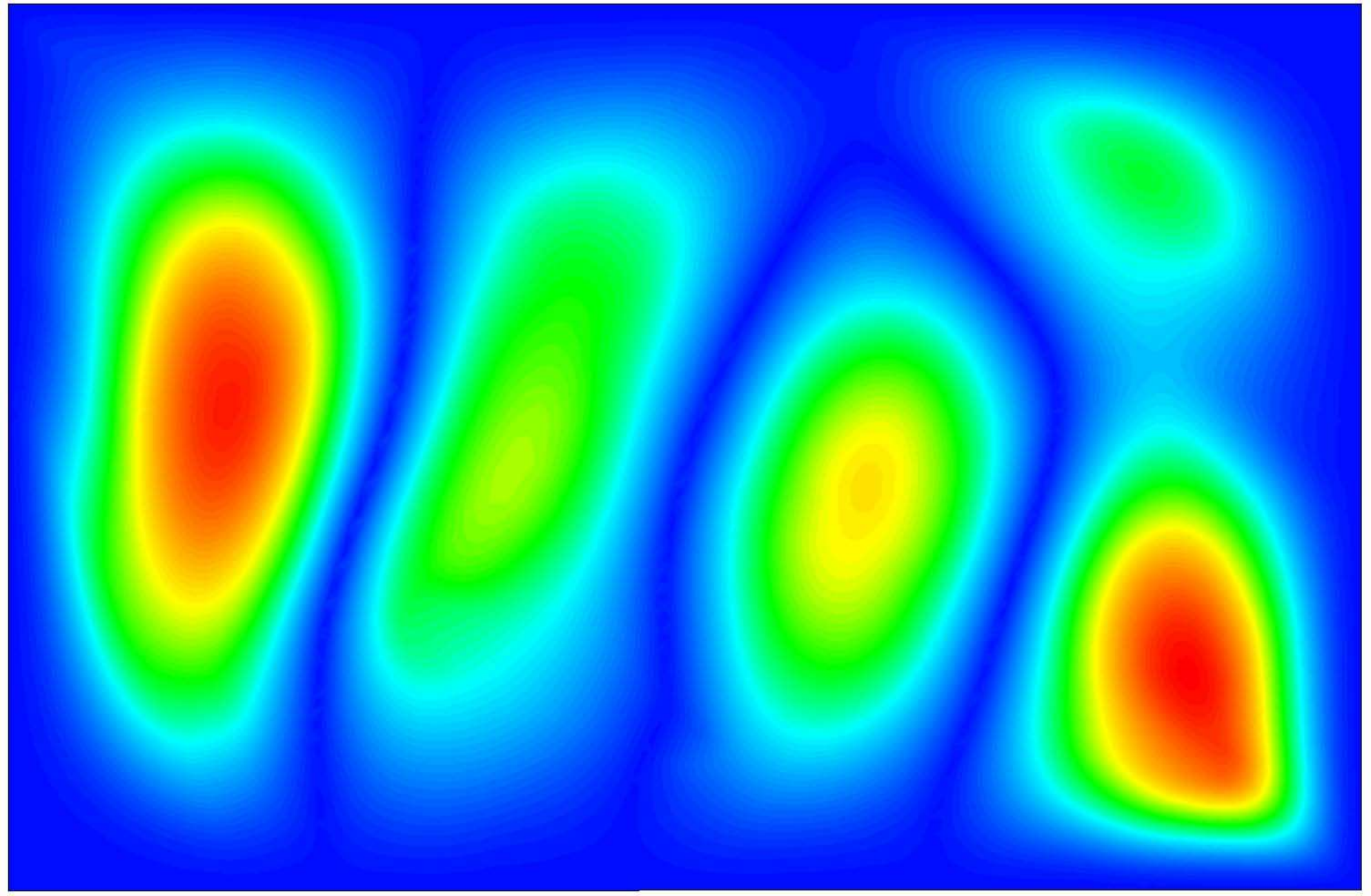}}&%
\includegraphics[width=0.24\linewidth]{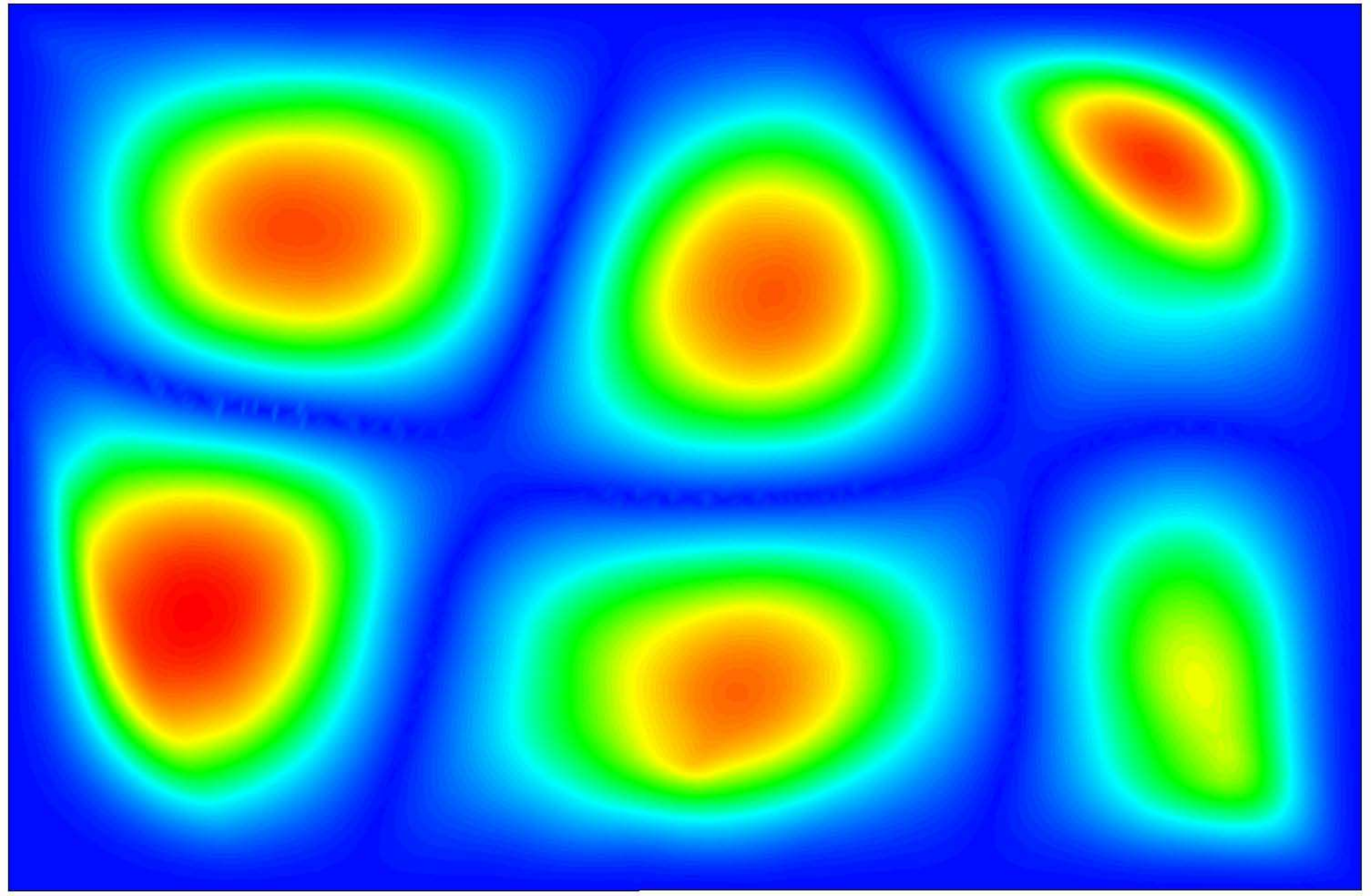}%
\vspace*{-.10cm}\\%
\multicolumn{2}{c}{\vspace*{-.15cm}%
271 Hz (Ns), 253 Hz (Ss), 234 Hz (ms)} &301 Hz (Ns), 279 Hz (Ss), 253 Hz (ms)\\%
\multicolumn{2}{c}{\vspace*{-.15cm}Mode (4,1)} &Mode (3,2)\\%
\multicolumn{2}{c}{$\overbrace{\text{274 Hz\hspace{0.24\linewidth}295 Hz}}$} &303 Hz%
\vspace*{0.05cm}\\
\includegraphics[width=0.24\linewidth]{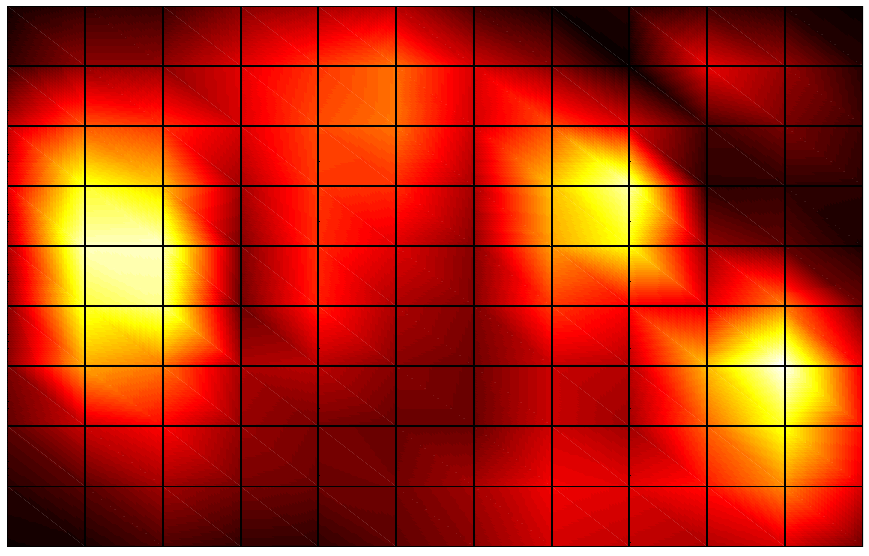}&%
\includegraphics[width=0.24\linewidth]{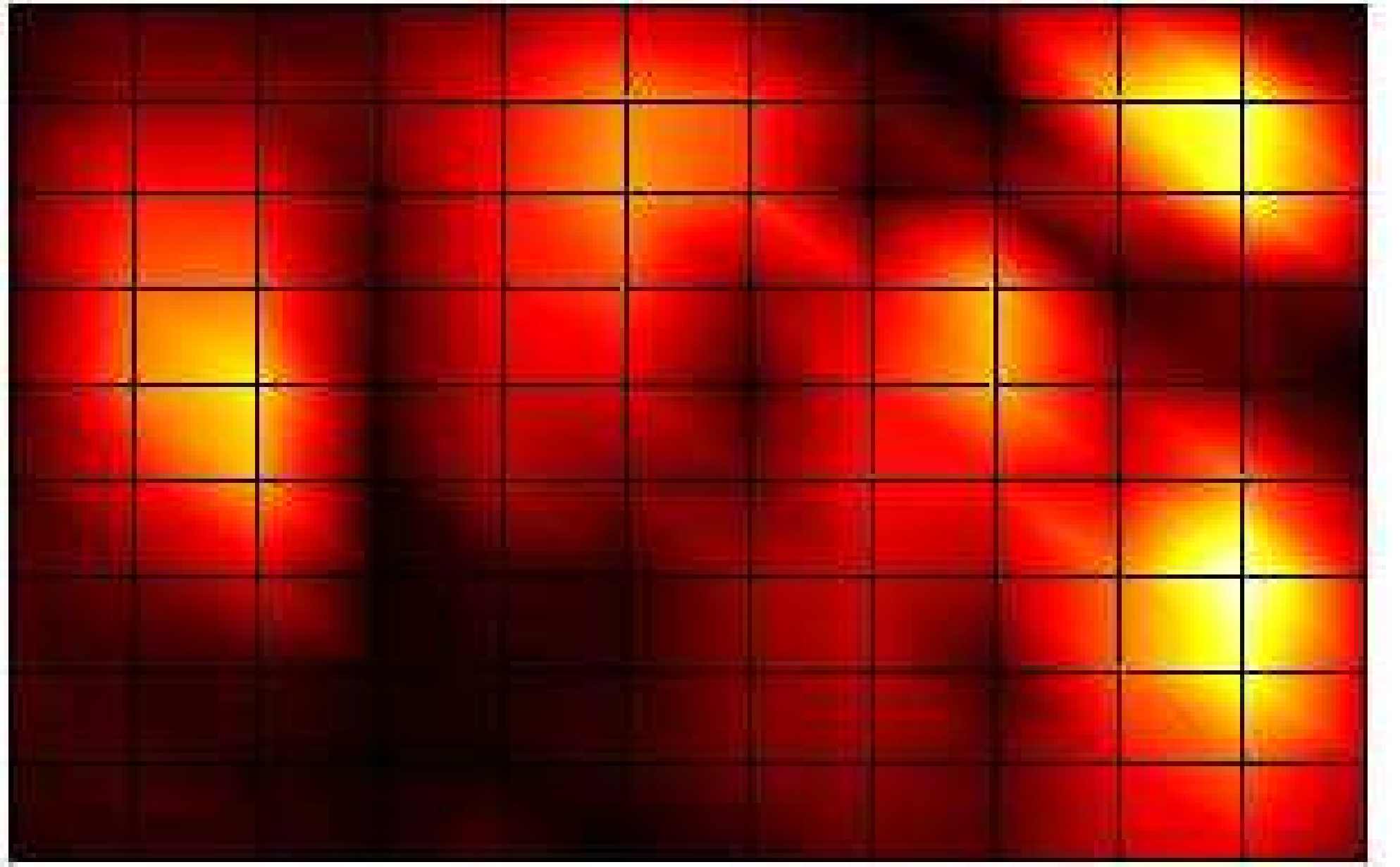}&%
\includegraphics[width=0.24\linewidth]{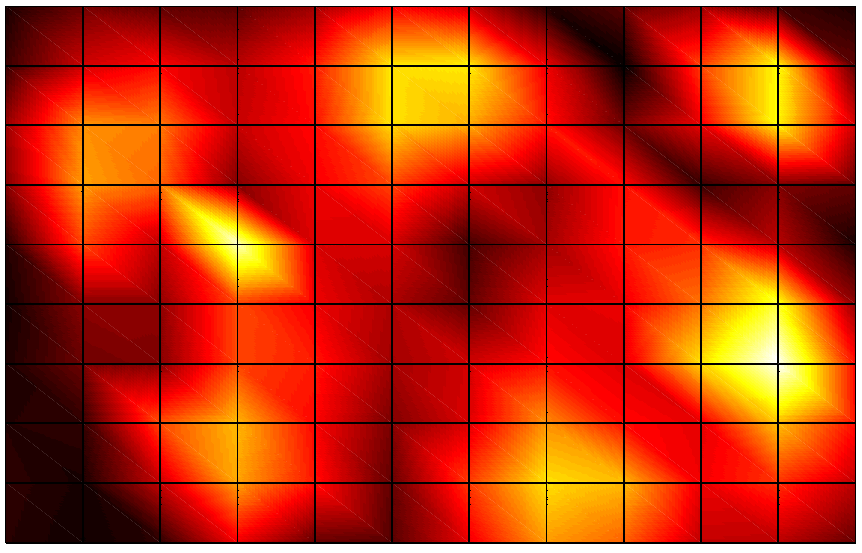}\\%
\end{tabular}
\end{center}
\caption[aaa]{First modal shapes and modal frequencies: numerical (upper lines) and experimental (lower lines). The modal frequencies of the numerical modes are reported for the Norway spruce (Ns), Sitka spruce (Ss) and mediocre spruce (ms), in this succession order. The modal shapes are those given with mediocre spruce. See text for the names and the grouping of modes.}
\label{fig:modes_castem}
\end{figure}

The first (up to 300 Hz) experimental and numerical modal shapes are presented in Fig.~\ref{fig:modes_castem}, together with the modal frequencies. As expected, the numerical modal frequencies (labeled "Ns", "Ss", "ms" when obtained with Norway spruce, Sitka spruce and mediocre spruce respectively) are sensitive to the elastic constants and density of wood. Without entering into a discussion of parameter or condition sensitivity which is not the focus of this article, one observes also that the numerical frequencies are little sensitive to the crown or to the string loading condition and that this sensitivity decreases with the modal order (Table~\ref{tab:LoadCrown}). These results are quite comparable, in relative magnitude, to the numerical results given by Mamou-Mani \emph{et al.}~\cite{MAM2008}: in this reference, see Fig.~3 for $b=1$ (the value that we have chosen for crowning, expressed in the units of this reference\footnote{We are skeptical on the practical character of crowning values corresponding to $b>2$.}) and Fig.~4, for a transverse displacement of 4~mm\footnote{We have reservations on the applicability of string loading that would reduce the initial crowning by more than one $b$ unit, here: 8~mm.}. However, one should consider all these results with care, due to a methodological problem: as explained in Sec.~\ref{sec:FEM}, residual stresses induced by manufacturing are ignored in our study, as well as in~\cite{MAM2008}. Since residual stresses are likely to induce changes in modal frequencies, comparable to the (small) changes due to string loading, the results given in Table~\ref{tab:LoadCrown}, as well as the predictions given by the model simulated in~\cite{MAM2008}, can only be considered as roughly indicative, with respect to crowning and string loading.

\begin{table}[ht!]
\begin{center}
\begin{tabular}{|c||c|c|c|c|c|c|c|c|c|c|c|}
\hline
Mode  &(1,1)&(2,1)&(3,1)&(1,2)&(2,2)&(4,1)&(3,2)\\
\hline

\hline
"C" and "L"&84&127&194&204&229&271&301\\
\hline
"C", no "L"&88&131&199&207&233&277&306\\
\hline
No "C", "L"&87&129&196&204&230&273&302\\
\hline
No "C", no "L"&88&131&200&209&234&280&309\\
\hline
\end{tabular}
\end{center} 
\caption[aaa]{First numerical modal frequencies, in Hz. Variations with the crown ("C") and loading ("L") conditions, with values given in Table~\ref{tab:caracmeca_num}, for Norway spruce.}
\label{tab:LoadCrown}
\end{table}

The modal shapes do not vary significantly for the three kinds of spruce that have been simulated (not shown here): this was expected, given the rather restricted variations of the ratio between $E\idr{L}$ and $E\idr{R}$ that have been allowed here. The same insensitivity to the loading and crown conditions (listed in Table~\ref{tab:LoadCrown}) has been observed.

The first mode which is observed experimentally, at 81~Hz, has no numerical counterpart. Looking at its modal shape, it appears that the boundary conditions cannot be considered as clamped or hinged all around the soundboard. According to visual inspection and therefore, approximately, the soundboard seems to be clamped or hinged along the top side, where the case is and which is more massive than the rim. Given the nodal line in the middle of the soundboard, the other boundary conditions at this particular frequency are dominated by inertia rather than elasticity. We therefore labelled this mode as "MMCM" (for mass-mass-clamped-mass). Since the boundary conditions at $f=0$~Hz are necessarily elastic (the piano stays at its place), it follows that the first resonance frequency of the boundary mobility, considered here as a whole for the sake of simplicity, is below 81 Hz. At least up to a hypothetical second resonance frequency, the inertial nature of boundary conditions have several consequences.

(a) The mobility of the boundary \emph{decreases} with frequency. Therefore, taking clamped (or hinged) boundary conditions becomes a better and better approximation, as frequency increases. Such constrained boundary conditions are generally assumed in the literature, as well as in our FEM simulations.

(b) Compared to clamped or hinged boundary conditions, the outer nodal line moves toward the \emph{inside} of the soundboard. Indeed, a hint of a nodal line can be seen at the bottom left corner of the (1,1) modal shape. However, the contrast between the mobility of the boundary and that of the soundboard is such that the outer nodal line cannot generally be distinguished from the boundary, to the exception of the MMCM and of the (1,1) modes.

(c) In comparison with clamped or hinged boundary conditions, inertial boundary conditions \emph{raise} the modal frequencies (that can also be understood by considering the inward shift of the outer nodal line, described above). Moreover, the relative shift in modal frequency is expected to decrease as frequency goes up (asymptotically, the modal frequencies do not depend on the boundary conditions). This is what is generally observed on the first modes depicted in Fig.~\ref{fig:modes_castem}: the experimental modal frequencies are systematically larger than their numerical counterparts. In consequence, choosing the characteristics of the soundboard material by fitting the first numerical modal frequencies to the experimental ones is not a good idea. The upper modal frequencies must be retained instead. This is why we consider that our soundboard is more probably cut in "mediocre spruce" than in either Sitka or Norway spruce (see Fig.~\ref{fig:densitemodale_guidedonde}).

(d) The mobility of the boundary is expected to be very low (see above) and to encounter erratic variations along its perimeter. Compared to strictly motionless boundary conditions (used in  FEM simulations), this is usually a cause for doubling some modes (twin modes) as observed for the (2,1), the (3,1) and the (4,1) modes which exhibit similar shapes but different modal frequencies. Modal \emph{families} of modes in low frequency have been observed on grand pianos, and attributed to the boundary conditions, by Suzuki~\cite{SUZ1986} in 1986 and Kindel~\emph{et~al.}~\cite{KIN1987} in 1987. Kindel~\emph{et~al.} observed up to three or four very similar modal shapes, differing mainly in the motion of the edge (the rim, also called the case) of the soundboard. Suzuki also noticed that mode splitting disappears when several bags of lead shot are put on the rim. He named such modes \emph{rim resonances}, somewhat misleadingly, in our opinion.

The labelling of the mode (1,2) is somewhat arbitrary. The deformation of the numerical mode labelled "(2,2)" is mainly located in the cut-off corner and we did not observe any clear experimental correspondent to this mode. One reason might be that no accelerometer was put in the cut-off corners during the experiment.

\begin{figure}[!ht]
\begin{center}
\scriptsize
\begin{tabular}{ccc}
\includegraphics[width=0.32\linewidth]{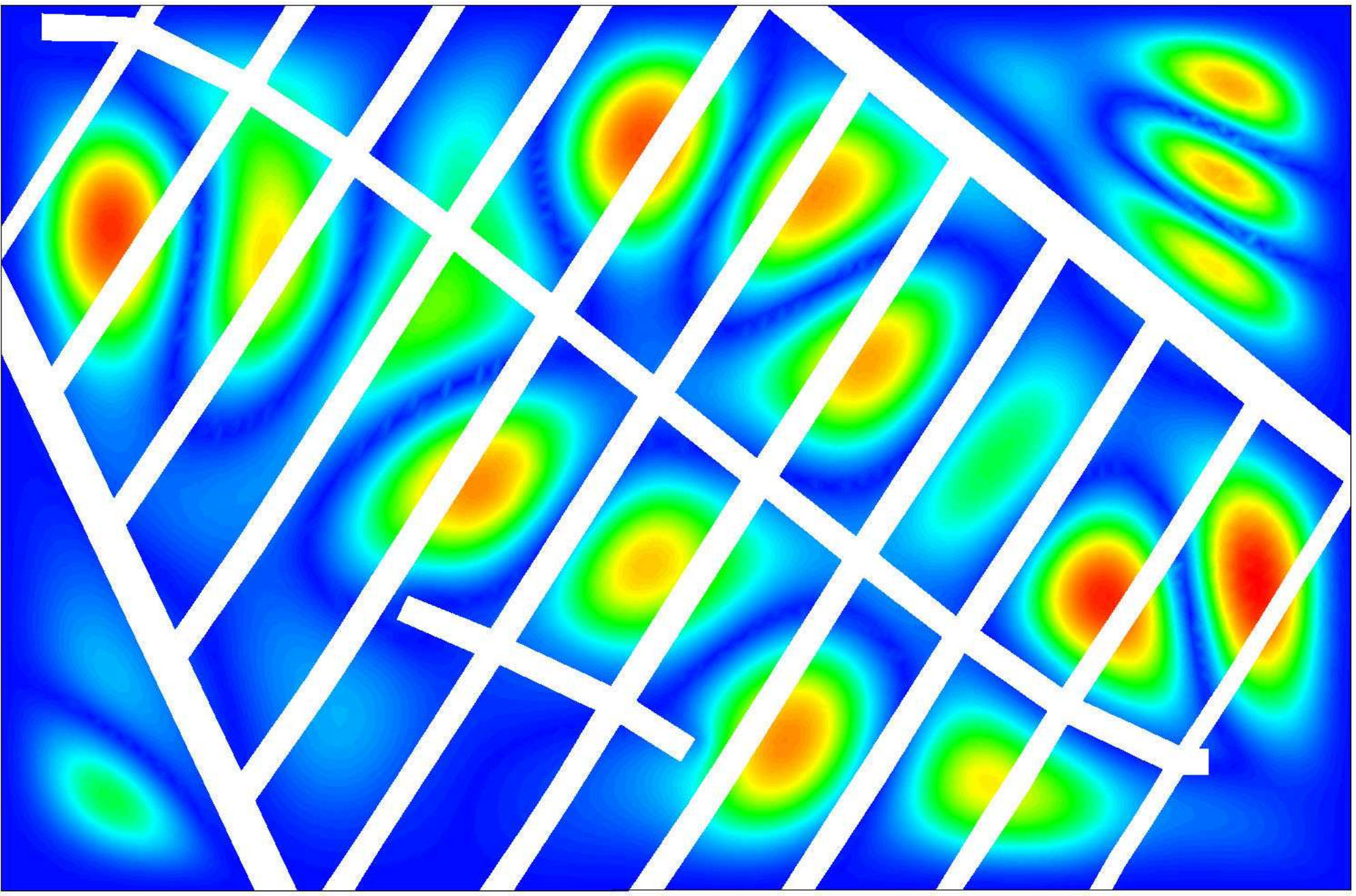}&%
\includegraphics[width=0.32\linewidth]{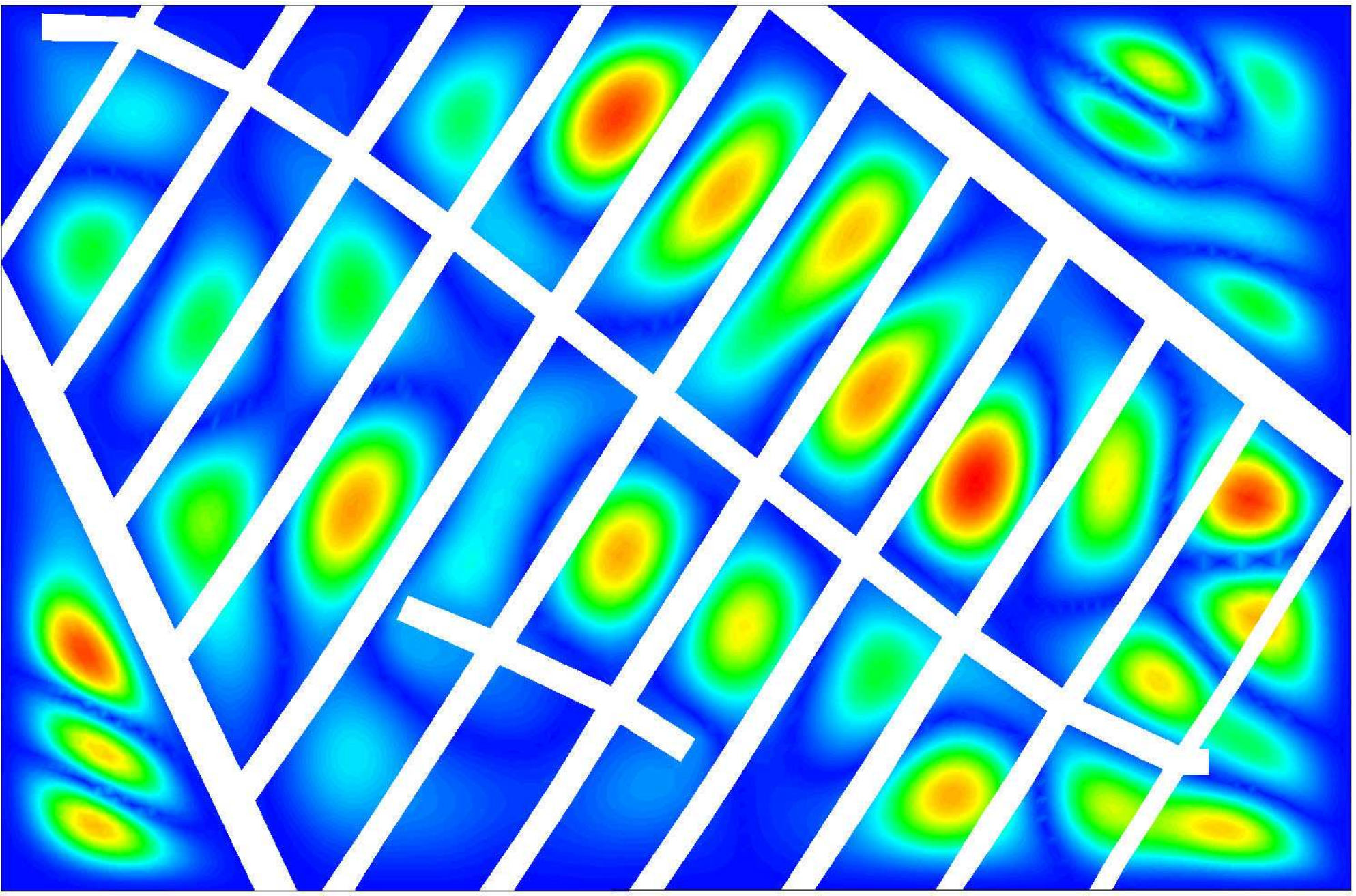}&%
\includegraphics[width=0.32\linewidth]{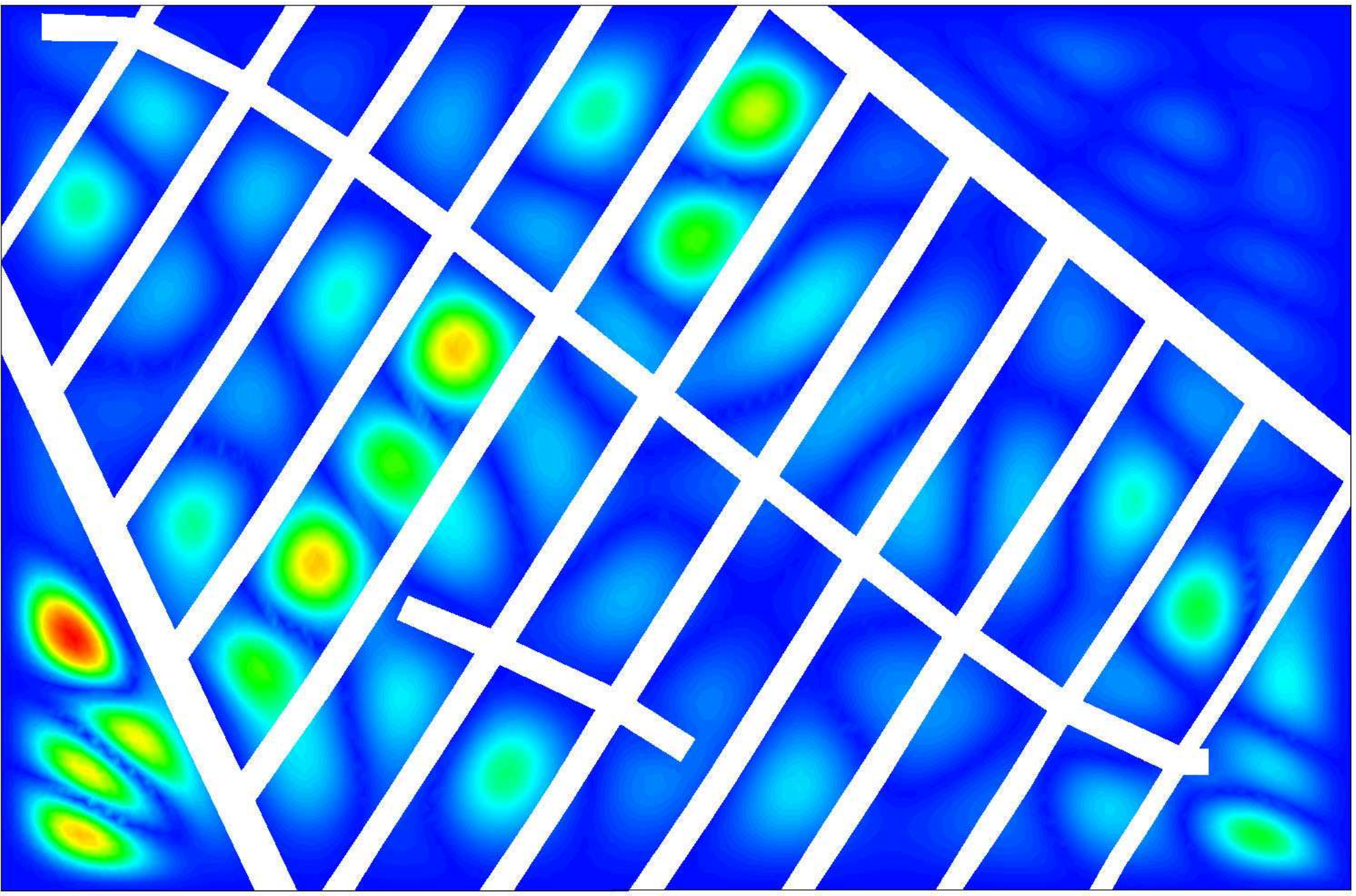}%
\vspace*{-.10cm}\\
\vspace*{-.15cm}%
Mode 28 -- 722 Hz &Mode 52 -- 1155 Hz &Mode 79 -- 1556 Hz\vspace*{0.5cm}\\
\includegraphics[width=0.32\linewidth]{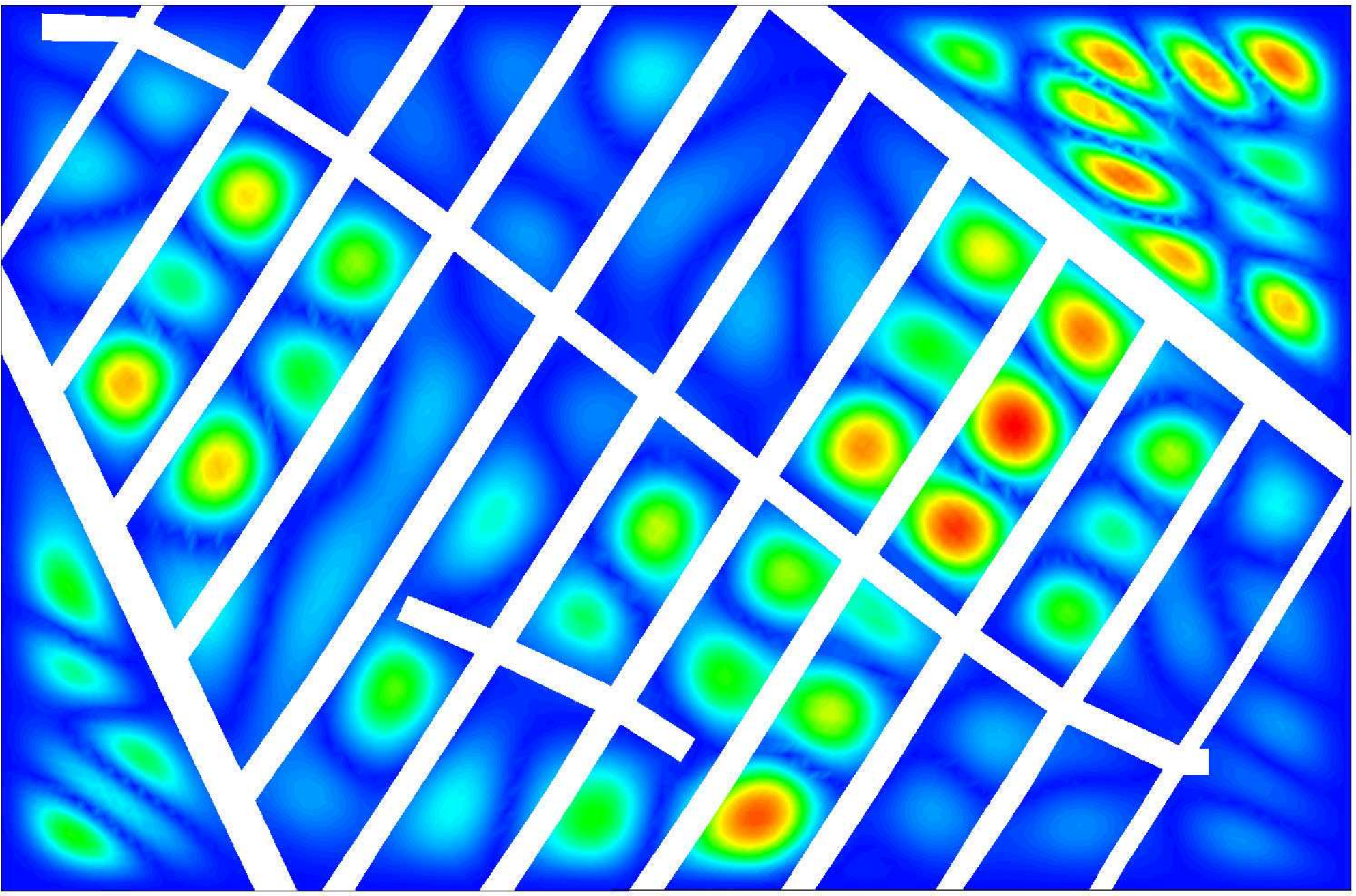}&%
\includegraphics[width=0.32\linewidth]{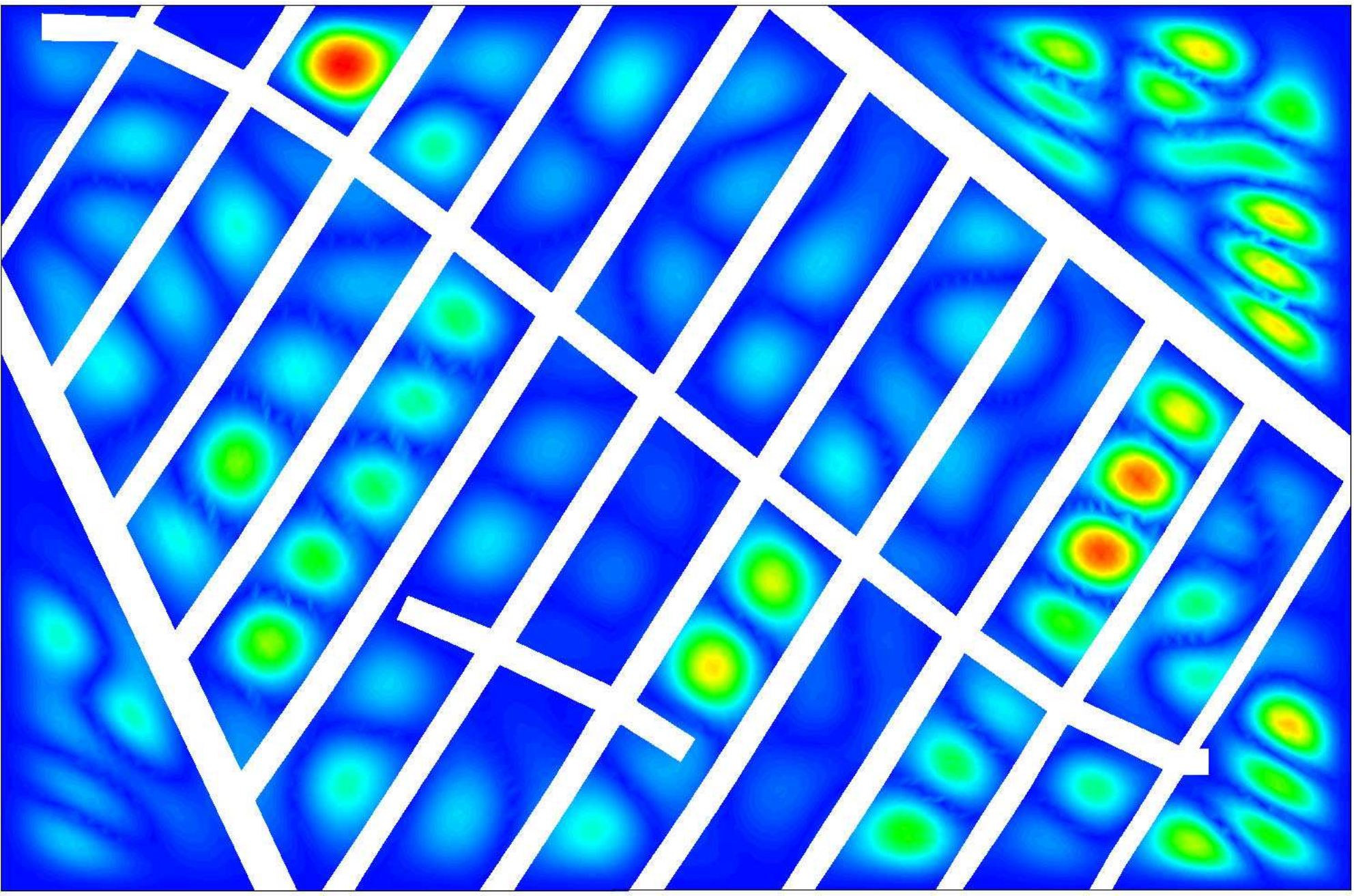}&%
\includegraphics[width=0.32\linewidth]{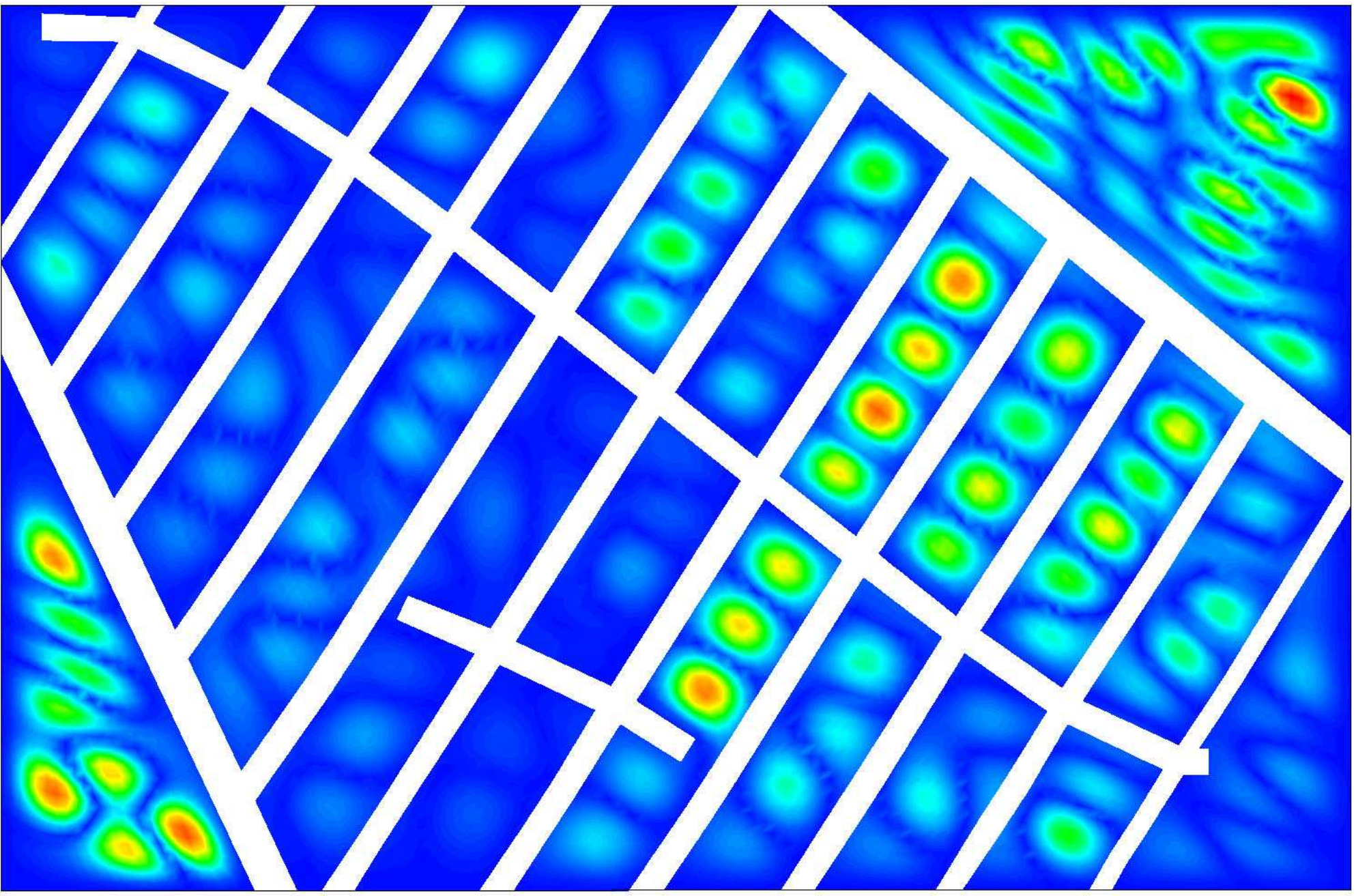}%
\vspace*{-.10cm}\\
\vspace*{-.15cm}%
Mode 93 -- 1728 Hz &Mode 119 -- 2103 Hz &Mode 167 -- 2733 Hz
\end{tabular}
\end{center}
\caption[aaa]{Examples of modal shapes obtained by the FEM analysis. Below \mbox{$\approx$ 1 kHz}, the ribs configuration is not apparent. Above \mbox{$\approx$ 1 kHz}, the vibration is confined between the ribs. In the ribbed zone of the soundboard and above that limit, most of the modes are localised in one or two areas, extending over a very few inter-rib spaces.}
\label{fig:castem_HF}
\end{figure}

Some of the higher modal shapes of the FEM are represented in Fig.~\ref{fig:castem_HF}, as computed with the characteristics of mediocre spruce. A transition occurs at \mbox{$\approx$ 1 kHz} (see also next section). Above this limit, the ribs confine wave propagation, as suggested by Nakamura~\cite{NAK1983}, observed by Moore \emph{et al.}~\cite{MOO2006} at 2837~Hz and characterised in terms of wave-numbers by Berthaut~\cite{BER2004}: the soundboard behaves like a set of more or less coupled structural wave-guides. Moreover, localisation seem to occur in this regime of vibration: whereas the modal shapes below 1~kHz extend throughout the entire wood panel, one can see that the modal shapes become localised above 1~kHz. It must be emphasised that, on plates, periodically spaced stiffeners restrict the possible values of the component of the wave-numbers in the direction normal to stiffeners (see \cite{MEA1996,MAC1980} for example). Localisation of the modes is almost certainly due to the non-uniform rib-spacing that is almost always observed in pianos (see Tab.~\ref{tab:espraidisseur} for the particular piano that has been investigated here), as in the case of the well-known Anderson localisation of waves in slightly disordered structures.

\subsection{Modal density}
\label{sec:density}
\begin{figure}[ht!]
\includegraphics[width=1\linewidth]{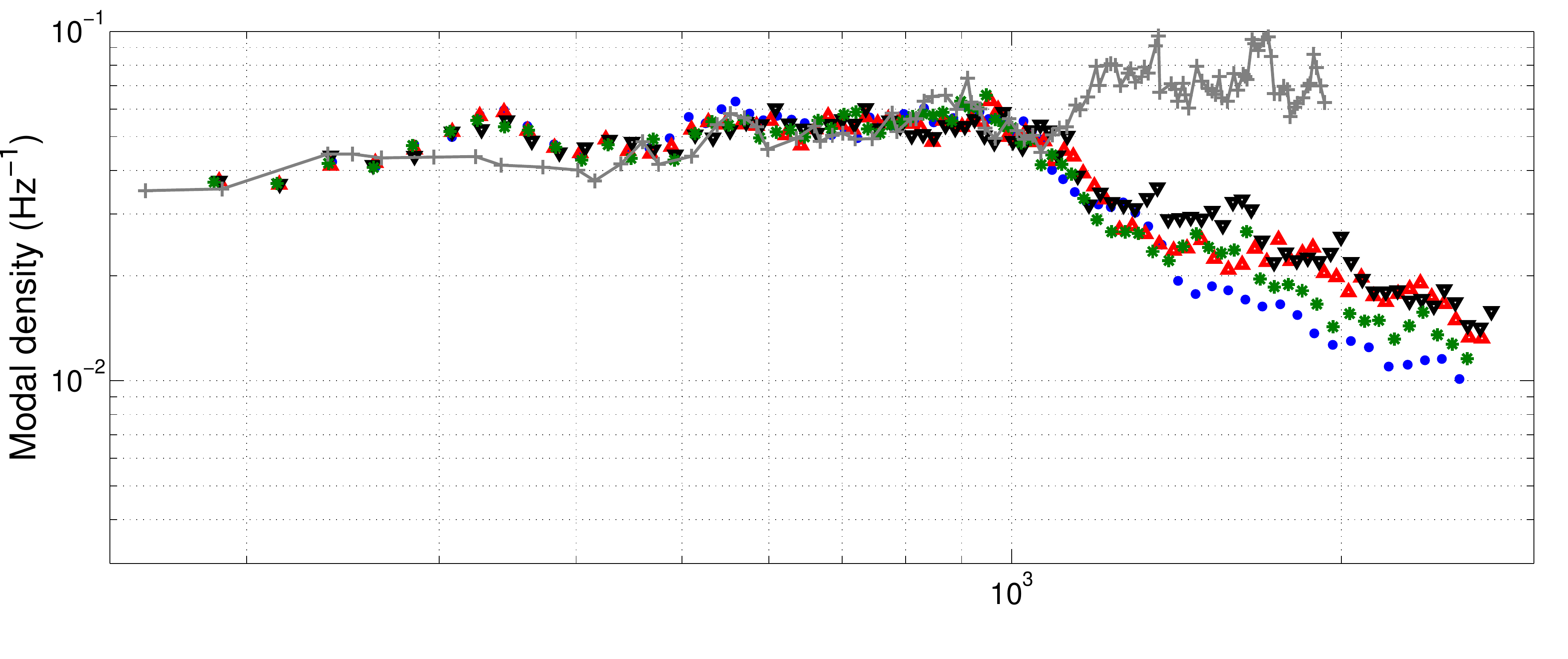}\\
\includegraphics[width=1\linewidth]{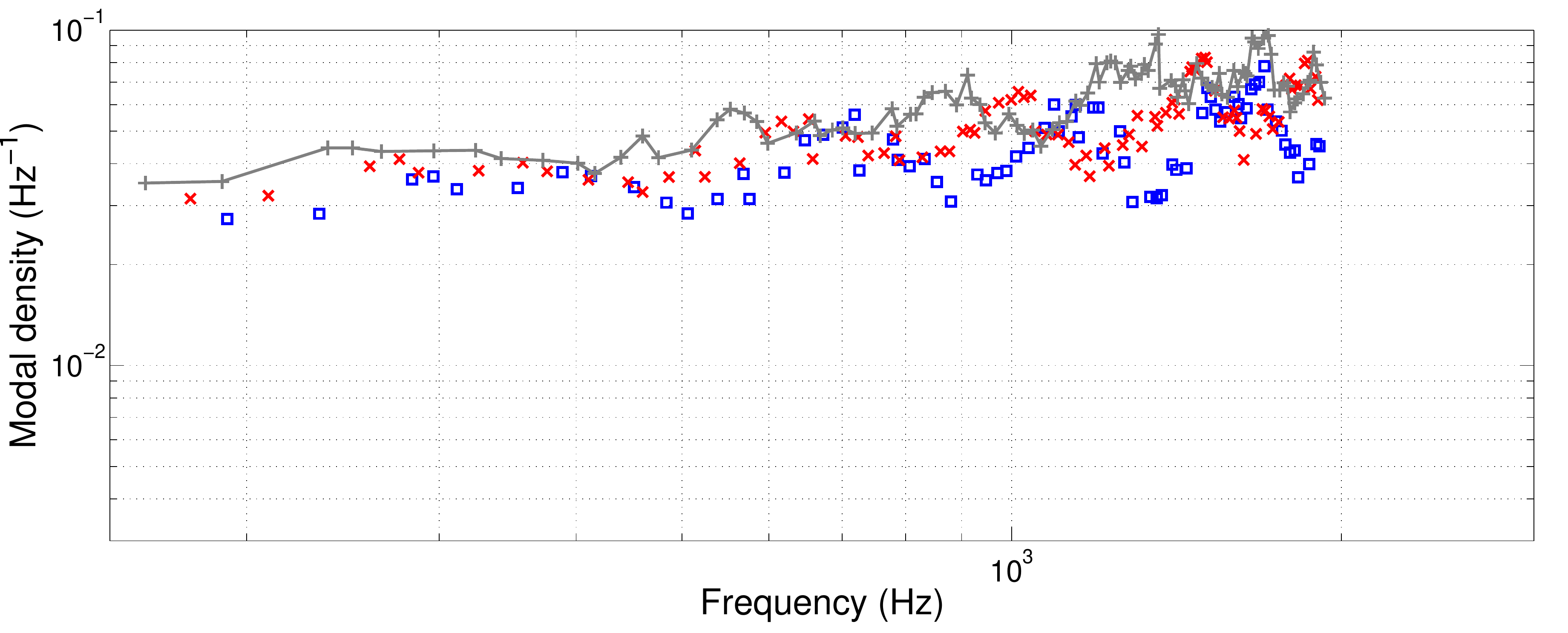}
\caption[aaa]{Estimations of the modal density in a piano soundboard. Each estimation is the reciprocal of the moving average on six successive modal spacings.\CR
Top frame: experimental determinations (apparent modal density) and numerical simulations with mediocre spruce (see below). Modes are measured up to 3~kHz. Dots: observed values at points \textbf{A}$_\mathbf{1}$ ({\color[rgb]{0,0,1}\tiny{$\bullet$}}), \textbf{A}$_\mathbf{2}$ ({\color[rgb]{1,0,0}$\blacktriangle$}), \textbf{A}$_\mathbf{3}$ ($\blacktriangledown$), and \textbf{A}$_\mathbf{5}$ ({\color[rgb]{0,0.5,0}\scriptsize{$\ast$}}), whose locations are given in Fig.~\ref{fig:table_exp_maillage}. {\color[rgb]{.5,.5,.5}$\bm{+}$\hdashrule[0.5ex]{2em}{1pt}{}$\bm{+}$\hdashrule[0.5ex]{2em}{1pt}{}$\bm{+}$}~: numerical modes given by FEM with the wood characteristics of mediocre spruce.\CR
Bottom frame: numerical modes given by FEM with the wood characteristics of Norway spruce ({\color[rgb]{0,0,1}$\bm{\Box}$}), Sitka spruce ({\color[rgb]{1,0,0}$\bm{\times}$}) and mediocre spruce ({\color[rgb]{.5,.5,.5}$\bm{+}$\hdashrule[0.5ex]{2em}{1pt}{}$\bm{+}$\hdashrule[0.5ex]{2em}{1pt}{}$\bm{+}$}).
}
\label{fig:densitemodale_guidedonde}
\end{figure}

Estimations of the modal density of the soundboard are represented in Fig.~\ref{fig:densitemodale_guidedonde} up to 3 kHz. They have been obtained as the reciprocal of the moving average on six successive estimated modal spacings. In the top frame of Fig.~\ref{fig:densitemodale_guidedonde}, the estimation is done independently at four points of measurement (see Fig.~\ref{fig:table_exp_maillage} for the exact locations): the estimated quantity is the \emph{apparent} modal density at that point. The average modal spacing (inverse of the modal density) is around 22~Hz for the $21$ lowest modes, in agreement with comparable low-frequency studies ($\approx 25$~Hz for a similar upright piano in ref.~\cite{DER1997} and $\approx 22$~Hz for a baby grand in ref.~\cite{SUZ1986}).

Although a parametric study is not in the scope of this paper, we present also the estimation of the modal density for numerical modes obtained with several sets of wood characteristics (see Table~\ref{tab:caracmeca_num}), in the bottom frame of Fig.~\ref{fig:densitemodale_guidedonde}. The case of mediocre spruce is also reported in the top frame. The frequency evolution $n(f)$ of the modal density reveals two distinct vibratory regimes of the structure.

Below $1.1$~kHz, the four experimental sets of results are almost similar. This means that the modes which are detected at each measurement point extend over the whole soundboard, as confirmed by the experimental and the numerical modal shapes that are shown in Fig.~\ref{fig:modes_castem}. The modal density increases slowly and tends towards a constant value of about $0.06~\text{modes}~\text{Hz}^{-1}$: the soundboard seems to behave more or less like a homogeneous plate.  Considering the highest modes in this frequency domain (as explained in the previous section), it appears that the simulations with mediocre spruce are those which best fit the experimental data.

The slow rise of $n(f)$ with frequency is characteristic of \emph{constrained} boundary conditions~\cite{XIE2004}. This experimental observation combined with the analysis of the lowest modes (see previous section) and with the observation of the mounting of the soundboard at its rim lead us to propose the following simplified scheme for the boundary conditions: the rotational degrees of freedom are blocked and the translational degrees of freedom are massive; as frequency increases, this scheme becomes equivalent to clamped boundary conditions.

For frequencies above $1.1$~kHz, $n(f)$, as measured at a given point, decreases significantly. Also, $n(f)$ is slightly but consistently different at each of the measured locations of the soundboard. In this frequency domain, the FEM and the experimental estimations of the modal density differ completely: the apparent modal density, estimated at one given point, is roughly the same everywhere but not the same as the global modal density given by a numerical simulation. This can be explained by the localisation of the vibrations, as suggested by Fig.~\ref{fig:castem_HF} and by the discussion in Sec.~\ref{sec:shapes}. The modal density given by the FEM takes into account \emph{all} the modes of the structure whereas the modes detected by one particular accelerometer are \emph{only} those having a significant level where the accelerometer is located. This also explains why the modal densities estimated at different locations are different. For example, \textbf{A}$_\mathbf{1}$ and \textbf{A}$_\mathbf{5}$ are near the corners of the soundboard (see Fig.~\ref{fig:table_exp_maillage}); they belong to shorter structural waveguides and "see" less modes than \textbf{A}$_\mathbf{2}$ and \textbf{A}$_\mathbf{3}$ which are located near the centre of the board where the waveguides are longer.

%% file: Appendix_MR.tex
Volterra series \cite{Hasler1999} are a means to express the relationship between the input $e(t)$ and the output $s(t)$ of any weakly nonlinear system~\cite{Palm1978,Boyd1985} as a series of multiple convolution integrals:

		\begin{equation}
		\label{eq:Volterra}
	{\textstyle  	s(t) =  }\sum_{k=1}^{+\infty} \int_0^{+\infty} \hspace{-0.3cm} {\scriptstyle \ldots} \int_0^{+\infty}{\textstyle  v_k(\tau_1,{\scriptstyle \ldots},\tau_k)e(t-\tau_1) {\scriptstyle \ldots} e (t-\tau_k) \dd\tau_1 {\scriptstyle \ldots} \dd\tau_k}
	\end{equation}
	
	The functions $\{v_k(\tau_1,{\scriptscriptstyle \ldots},\tau_k)\}_{k \in \mathbb{N}^*}$ are called Volterra kernels and completely characterise the nonlinear system under study. Volterra series are a generalisation of the simple convolution operator used for linear systems. Following the idea of \emph{linear transfer functions} for linear systems, \emph{nonlinear transfer functions} can be obtained by expressing the Volterra kernels in the frequency domain \emph{via} a multidimensional Fourier transform:
	\begin{equation}
		\label{eq:VolterraF}
		\forall k \in \mathbb{N}^* \quad
		V_k(f_1,{\scriptstyle \ldots},f_k) = \int_{\mathbb{R}_{+}^{k}} v_k(\tau_1,{\scriptscriptstyle \ldots},\tau_k) e^{-i2 \pi f_1 \tau_1} \ldots e^{-i2 \pi f_k \tau_k} \dd\tau_1 {\scriptstyle \ldots} \dd\tau_k
	\end{equation}

The association of two weakly nonlinear systems $\mathcal{S}_F$ and $\mathcal{S}_G$ is now considered, as shown in Fig.~\ref{fig:NL_chain}. The families of kernels $\{F_k(f_1,{\scriptscriptstyle \ldots},f_k)\}_{k \in \mathbb{N}^*}$, $\{G_k(f_1,{\scriptscriptstyle \ldots},f_k)\}_{k \in \mathbb{N}^*}$ and, $\{H_k(f_1,{\scriptscriptstyle \ldots}, f_k)\}_{k \in \mathbb{N}^*}$ fully represent the systems $\mathcal{S}_F$, $\mathcal{S}_G$ and, $\mathcal{S}_H$. In the present case, the kernels of $\mathcal{S}_H$ can be expressed as functions of the kernels of $\mathcal{S}_F$ and $\mathcal{S}_G$ as follows \cite{Hasler1999}:
	
	\begin{equation}
	\begin{array}{ll}
										H_k(f_1,{\scriptscriptstyle \ldots},f_k) =  
										\displaystyle\sum_{p=1}^{k}
										\sum_{
											\mathcal{M}_p^k
										}
										&
										F_{m_1}(f_1,{\scriptscriptstyle \ldots}, f_{m_1}) 
										\times \ldots  
										\vspace{-0.4 cm}
										\\
										&
										\ldots \times 
										F_{m_p}(f_{m_1+ \ldots + m_{p-1}+1},{\scriptscriptstyle \ldots}, f_k)
										\\
										&
										\times  
										G_{p}(f_1+ \ldots + f_{m_1}, {\scriptscriptstyle \ldots},f_{m_1+ \ldots + m_{p-1}+1} + f_n)
										\vspace{0.4 cm}  
										\\
										&
										\forall k \in \mathbb{N}^* \quad 
										\text{and with}
										\quad \mathcal{M}_p^k = 
											\left\{
											\begin{array}{c} 
												\scriptstyle{m_1+ \ldots + m_p = k} \\
												\scriptstyle{m_1, \,\ldots\,, m_p \geq 1}
											\end{array}	
											\right. 
	\end{array}
	\label{eq:serieNL}
	\end{equation}
	
	For the two first terms, Eq.~\eqref{eq:serieNL} reduces to:
	
	\begin{equation}
	\left\{
	\begin{array}{ll}
	H_1(f_1) &= \ F_1(f_1)G_1(f_1)
	\vspace{0.1 cm} \\
	H_2(f_1,f_2) &= \ F_2(f_1,f_2)G_1(f_1+f_2) + F_1(f_1)F_1(f_2)G_2(f_1,f_2)
	\end{array}
	\right.
	\end{equation}
	
	This proves the intuitive results that the \emph{linear transfer function} describing the linear behaviour of a weakly nonlinear system $\mathcal{S}_G$ following another weakly nonlinear system $\mathcal{S}_F$ is simply the product of the \emph{linear transfer functions} of those two systems.

%% file: NL_Method.tex
	The mathematical foundations of the method used for the estimation of the elements of a cascade of Hammerstein models~\cite{REB2011,NOV2010} are given in this section. In such a system, each branch is composed of one nonlinear static polynomial element followed by a linear one $\beta_n(t)$ and the relation between its input $e(t)$ and its output $s(t)$ is given by Eq.~\eqref{eq:hamm}, where $*$ denotes the convolution.
	
	\begin{equation} s(t) = \sum_{n=1}^{N}\beta_n*e^n(t) \label{eq:hamm} \end{equation}
		
		To experimentally cover the frequency range on which the system under study is to be identified, cosines with time-varying frequencies are used. If \mbox{$e(t)=\cos[\Phi(t)]$} is the input of the cascade of Hammerstein models, the output of the nonlinear block $e^i(t)$ is rewritten using Chebyshev polynomials as in Eq.~\eqref{eq:tchebi_ici}. Details of the computation of the Chebyshev matrix $\boldsymbol C$ are provided in~\cite{REB2011,NOV2010}.  
	
	\begin{equation} \forall i \in \{1 \ldots N\} \quad e^i(t) =\cos^i[\Phi(t)] = \sum_{k=0}^i C(i,k) \cos[k \Phi(t)] \label{eq:tchebi_ici} \end{equation}
	
	 When the instantaneous frequency of $e(t)$ is increasing exponentially (from $f_1$ to $f_2$ in a time interval $T$), the signal is called ``exponential sine sweep''. It can be shown~\cite{REB2011,NOV2010} that by choosing \mbox{$T_m=\left(2 m \pi - \dfrac{\pi}{2}\right)\dfrac{\ln{f_2/f_1}}{2\pi f_1}$} with $m \in \mathbb{N}^*$, one obtains:
	\begin{equation} 
 \label{eq:phaseProp}
	\forall k \in \mathbb{N^*}, \quad \cos(k \Phi(t)) = \cos( \Phi(t+ \Delta t_k))   
	 \quad \text{with} \quad \Delta t_k = \frac{T_m \ln{k}} {\ln{(f_2/f_1)}}  
	\end{equation}
which represents another expression of the $k^{th}$ term in the linearisation presented in Eq.~\eqref{eq:tchebi_ici}. 
			
		For any $T_m$-long exponential sine sweep, multiplying the phase by a factor $k$ yields the same signal, advanced in time by $\Delta t_k$. Using Eqs.~\eqref{eq:phaseProp} and \eqref{eq:hamm}, one obtains:
	
	\begin{equation} 
	s(t) = \sum_{n=1}^{N} \gamma_n*e(t+\Delta t_n) \quad \text{with} \quad \gamma_n(t) = \sum_{k=1}^{N} C(k,n) h_k(t)
	\label{eq:rel_ES} 
	\end{equation}
		
		$\gamma_n(t)$ corresponds to the contribution of the different kernels to the $n^{th}$ harmonic. In order to separately identify each kernel of the cascade of Hammerstein models, a signal $\tilde{e}(t)$, operating as an inverse of $e(t)$ in the convolution sense, is needed. The Fourier transform  $\tilde{E}(f)$ of the inverse filter $\tilde{e}(t)$ can be built in the frequency domain by means of Eq.~\eqref{eq:def_inv}, where $E^*(f)$ is the complex conjugate of $E(f)$, the Fourier transform of $e(t)$.
	
	\begin{equation}
	\tilde{E}(f) = \frac{1}{E(f)} \mathds{1}_{[-f_2,-f_1]\cap [f_1,f_2]}(f) \simeq \frac{E^{*}(f)}{|E(f)|^2 + \epsilon(f)}
	\label{eq:def_inv}
	\end{equation}

	$\epsilon(f)$ is a frequency-dependent real parameter chosen to be $0$ in the bandwidth and to have a large value outside of the bandwidth, with a continuous transition between the two domains. After convolving the output of the cascade of Hammerstein models $s(t)$ given in Eq.~\eqref{eq:rel_ES} with $y(t)$, one obtains: 
		
		\begin{equation} \tilde{e}*s(t) = \sum_{i=1}^{N} \gamma_i(t + \Delta t_n) \label{eq:final} \end{equation}

		Because $\Delta t_n \propto \ln(n)$ and $f_2>f_1$, the higher the order of linearity $n$, the more advanced is the corresponding $\gamma_n(t)$. Thus, if $T_m$ is chosen long enough, the different $\gamma_n(t)$ do not overlap in time and can be separated by simply windowing them in the time domain. Using Eq.~\eqref{eq:retour}, the family $\{\beta_n(t)\}_{n \in [1,N]}$ of the kernels of the cascade of Hammerstein models under study can then be fully extracted. 
		
		\begin{equation} 
				\begin{pmatrix}
					\beta_1(t)\\
					\vdots\\
					\beta_N(t)\\
				\end{pmatrix} 
				 = \boldsymbol A_c^T
				 \begin{pmatrix}
					\gamma_1(t)\\
					\vdots\\
					\gamma_N(t)\\
				\end{pmatrix} 
				\label{eq:retour}
		\end{equation} 
		
$(.)^T$ stands for matrix transposition and $\boldsymbol A_c$ is the Chebyshev matrix $\boldsymbol C$ defined earlier, from which the first column and the first row have been removed.

%% file: appendix_ribsV32.tex
\begin{figure}[ht!]
\begin{center}
\includegraphics[width=0.55\linewidth]{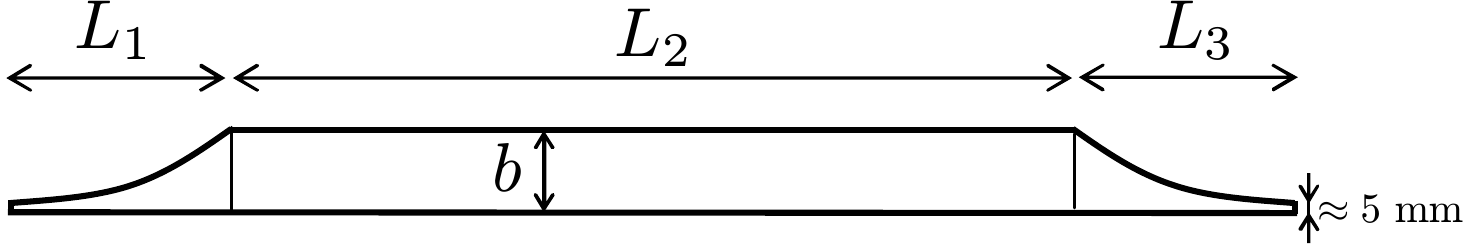}\hspace{0.05\linewidth}%
\includegraphics[width=0.34\linewidth]{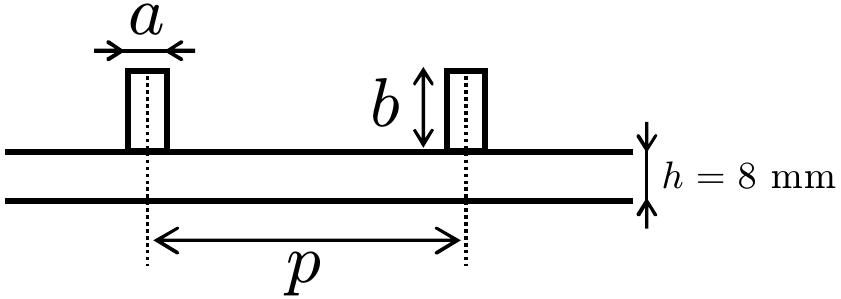}\\
\hspace*{.1\linewidth}(a)\hspace{.45\linewidth}(b)
\end{center}
\caption{Rib and soundboard geometry. (a) Side view of a rib. (b) Partial front view of the ribbed soundboard.}
 \label{fig:dimraidisseur}
\end{figure}

In order to ease the implementation of the FEM, the geometry of the ribs has been simplified: each rib is given a uniform height (or thickness) all along its length. In reality, ribs are tapered (Fig.~\ref{fig:dimraidisseur}), thus giving less stiffness to the soundboard near its edges. In the FEM, the rib thickness is averaged over the rib length, thus keeping the same rib mass but not exactly the same rigidity since this mechanical property is proportional to $h^3$. 
The retained geometry is given in Tabs.~\ref{tab:dimraidisseur} and \ref{tab:espraidisseur}. 

\vspace*{\baselineskip}

\begin{table}[!ht]
\begin{center}
\begin{tabular}{|c||c|c|c|c|c|c|c|c|c|c|c|}
\hline
Rib \#& 1 & 2 & 3 & 4 & 5 & 6 & 7 & 8 & 9 & 10 & 11\\
\hline

\hline
$L_1$& 16.0 & 18.5 & 17.5 & 16.5 & 16.5 & 12.0 & 13.0 & 15.0 & 15.0 & 15.0 & 10.0\\
\hline
$L_2$& 8.0& 19.0 & 33.5 & 48.5 & 49.0& 71.5 &59.5 &46.0 &37.0 &21.0 &11.0\\
\hline
$L_3$& 6.0 & 8.0 & 11.0& 13.0& 15.0& 15.0& 16.0& 17.0& 16.5& 22.5& 26.5\\
\hline
$a$  &  2.0 &   2.6   & 2.8   & 2.8  &  3.0  &  3.2   & 3.0 &   3.0 &   2.8   & 2.5   & 2.0\\
\hline
$b$&  2.3& 2.3& 2.6& 2.6& 2.6& 2.6& 2.6 & 2.6& 2.5& 1.8& 1.7\\
\hline
\hline
$b_\text{av}$&  1.6 &   1.8   & 2.1   & 2.2  &  2.2  & 2.3 & 2.3   & 2.2 &   2 &   1.4  & 1.2\\
\hline
\end{tabular}
\end{center}
\caption{Dimensions of the ribs. Rib \#1 is at the treble end of the soundboard. $b_\text{av}$ is the averaged thickness of the rib. All values are in cm.}
\label{tab:dimraidisseur}
\end{table}

\vspace*{\baselineskip}

\begin{table}[!hb]
\begin{center}
\begin{tabular}{|c||c|c|c|c|c|c|c|c|c|c|}
\hline
Rib \#& 1-2 & 2-3 & 3-4 & 4-5 & 5-6 & 6-7 & 7-8 & 8-9 & 9-10 & 10-11\\
\hline

\hline
$p$ & 12.7 & 12.8&12.9 &13.5& 13.0 &13.7& 13.0& 13.0& 12.7& 12.8\\
\hline
\end{tabular}
\end{center}
\caption{Inter-rib space $p$ (in cm). Note that the spacing does not seem to follow a simple law and may simply be considered as irregular.}
\label{tab:espraidisseur}
\end{table}

%% file: Vibrations_JSV_V4.bbl
\begin{thebibliography}{10}
\expandafter\ifx\csname url\endcsname\relax
  \def\url#1{\texttt{#1}}\fi
\expandafter\ifx\csname urlprefix\endcsname\relax\def\urlprefix{URL }\fi

\bibitem{EGE2009_2bib}
K.~Ege, La table d'harmonie du piano -- {É}tudes modales en basses et moyennes
  fréquences ({T}he piano soundboard -- {M}odal studies in the low- and
  mid-frequency ranges), ch.~1 (in {E}nglish), sec.~2, Ph{D} thesis, École
  polytechnique, Palaiseau, France (2009).
\newline\urlprefix\url{http://tel.archives-ouvertes.fr/docs/00/46/17/77/PDF/These\_Ege.pdf}

\bibitem{EGE2009}
K.~Ege, X.~Boutillon, B.~David, High-resolution modal analysis, Journal of
  Sound and Vibration 325~(4-5) (2009) 852--869.

\bibitem{TOU2002}
C.~Touzé, O.~Thomas, A.~Chaigne, Asymmetric non-linear forced vibrations of
  free-edge circular plates. part 1: Theory, Journal of Sound and Vibration
  258~(4) (2002) 649--676.

\bibitem{ASK1992}
A.~Askenfelt, E.~V. Jansson, On vibration sensation and finger touch in
  stringed instrument playing, Music Perception 9~(3) (1992) 311--349.

\bibitem{HUN1978}
T.~C. Hundley, H.~Benioff, D.~W. Martin, Factors contributing to multiple rate
  of piano tone decay, Journal of the Acoustical Society of America 64~(5)
  (1978) 1303--1309.

\bibitem{Palm1978}
G.~Palm, On representation and approximation of nonlinear systems, Biological
  Cybernetics 31~(2) (1978) 119--124.

\bibitem{Boyd1985}
S.~Boyd, L.~O. Chua, Fading memory and the problem of approximating nonlinear
  operators with {Volterra} series, IEEE Transactions on Circuits and Systems
  32~(11) (1985) 1150--1161.

\bibitem{REB2011}
M.~Rébillat, R.~Hennequin, E.~Corteel, B.~F. Katz, Identification of cascade of
  {H}ammerstein models for the description of nonlinearities in vibrating
  devices, Journal of Sound and Vibration 330~(5) (2011) 1018--1038.

\bibitem{NOV2010}
A.~Novak, L.~Simon, F.~Kadlec, P.~Lotton, Nonlinear system identification using
  exponential swept-sine signal, IEEE Transactions On Instrumentation and
  Measurement 59~(8) (2010) 2220--2229.

\bibitem{ROY1989}
R.~Roy, T.~Kailath, Esprit - estimation of signal parameters via rotational
  invariance techniques, IEEE Transactions on Acoustics Speech and Signal
  Processing 37~(7) (1989) 984--995.

\bibitem{BAD2006}
R.~Badeau, B.~David, G.~Richard, A new perturbation analysis for signal
  enumeration in rotational invariance techniques, IEEE Transactions on Signal
  Processing 54~(2) (2006) 450--458.

\bibitem{CASTEM}
Cast3{M} 2011, finite element software, http://www-cast3m.cea.fr/ (last viewed
  july 30, 2012).

\bibitem{CON1996_2}
H.~A. Conklin, Design and tone in the mechanoacoustic piano .part 2. piano
  structure, Journal of the Acoustical Society of America 100~(2) (1996)
  695--708.

\bibitem{HAI1979}
D.~W. Haines, On musical instrument wood, Catgut Acoustical Society Newsletter
  31 (1979) 23--32.

\bibitem{BER2004}
J.~Berthaut, Contribution à l'identification large bande des structures
  anisotropes. {A}pplication aux tables d'harmonie des pianos ({C}ontribution
  on broad-band identification of anisotropic structures. {A}pplication to
  piano soundboards), Ph{D} thesis, {É}cole centrale de Lyon, Lyon, France
  (2004).

\bibitem{SUZ1986}
H.~Suzuki, Vibration and sound radiation of a piano soundboard, Journal of the
  Acoustical Society of America 80~(6) (1986) 1573--1582.

\bibitem{DER1997}
P.~Dérogis, Analyse des vibrations et du rayonnement de la table d'harmonie
  d'un piano droit et conception d'un système de reproduction du champ
  acoustique ({A}nalysis of the vibro-acoustics behaviour of a piano soundboard
  and conception of a multiloudspeaker setup wich synthesis directivity
  patterns), Ph{D} thesis, Université du Maine, Le Mans, France (1997).

\bibitem{BER2003}
J.~Berthaut, M.~N. Ichchou, L.~Jezequel, Piano soundboard: structural behavior,
  numerical and experimental study in the modal range, Applied Acoustics
  64~(11) (2003) 1113--1136.

\bibitem{MAM2008}
A.~Mamou-Mani, J.~Frelat, C.~Besnainou, Numerical simulation of a piano
  soundboard under downbearing, Journal of the Acoustical Society of America
  123~(4) (2008) 2401--2406.

\bibitem{KIN1987}
J.~Kindel, I.-C. Wang, Modal analysis and finite element analysis of a piano
  soundboard, in: Proceedings of the 5th International Modal Analysis
  Conference, Union College, Schenectady, New York, 1987, pp. 1545--1549.

\bibitem{NAK1983}
I.~Nakamura, The vibrational character of the piano soundboard, in: Proceedings
  of the 11th ICA, Vol.~4, Paris, 1983, pp. 385--388.

\bibitem{MOO2006}
T.~R. Moore, S.~A. Zietlow, Interferometric studies of a piano soundboard,
  Journal of the Acoustical Society of America 119~(3) (2006) 1783--1793.

\bibitem{MEA1996}
D.~J. Mead, Wave propagation in continuous periodic structures: Research
  contributions from southampton, 1964-1995, Journal of Sound and Vibration
  190~(3) (1996) 495--524.

\bibitem{MAC1980}
B.~R. Mace, Periodically stiffened fluid-loaded plates, 2. {R}esponse to line
  and point forces, Journal of Sound and Vibration 73~(4) (1980) 487--504.

\bibitem{XIE2004}
G.~Xie, D.~J. Thompson, C.~J.~C. Jones, Mode count and modal density of
  structural systems: relationships with boundary conditions, Journal of Sound
  and Vibration 274~(3-5) (2004) 621--651.

\bibitem{Hasler1999}
M.~Hasler, Ph\'enom\`enes non lin\'eaires (nonlinear phenomenon), ch. 3:
  S\'eries de Volterra (Volterra series), \'Ecole Polytechnique F\'ed\'erale de
  Lausanne, 1999.

\end{thebibliography}
